\documentclass[12pt]{article}

\usepackage[utf8]{inputenc}
\usepackage{latexsym, graphicx} 
\usepackage{amsmath,amsfonts,amssymb}
\usepackage{slashed}
\usepackage{cancel}
\usepackage{mathrsfs}
\usepackage{tensor}
\usepackage{xcolor}
\usepackage{cite}
\usepackage[colorlinks=true,linkcolor=blue,citecolor=blue,urlcolor=blue,filecolor=black]{hyperref}
\usepackage{comment}
\usepackage{tikz,float}
\usepackage{tikz-3dplot}
\usetikzlibrary{arrows.meta}

\renewenvironment{thebibliography}[1]{
	\begin{oldthebibliography}{#1}
		\setlength{\itemsep}{2pt}
		\setlength{\parskip}{2pt}
	}
	{
	\end{oldthebibliography}
}

\numberwithin{equation}{section} 

\newcommand*{\scri}{\ensuremath{\mathscr{I}}}

\newcommand*{\x}{\mathbf{x}}

\newcommand{\ph}{\hat p}

\newcommand{\p}{\hat{p}}

\begin{document}
	
	\begin{titlepage}
		\thispagestyle{empty}
		
		\begin{flushright}
		\end{flushright}
		
		\vskip1cm
		
		\begin{center}  
			{\Large\textbf{Hilbert space in the flat limit of AdS/CFT\\
					\vskip 2mm II.~Massless particles}}
			
			\vskip1cm
			
			\centerline{Kevin Nguyen}
			
			\vskip1cm
			
			{\it{Universit\'e Libre de Bruxelles and International Solvay Institutes,\\ ULB-Campus Plaine
					CP231, 1050 Brussels, Belgium}}\\
			\vskip 1cm
			{kevin.nguyen2@ulb.be}
			
		\end{center}
		
		\vskip1cm
		
		\begin{abstract}
			Building on Part I of this series, where we described the emergence of massive scattering states, we demonstrate how unitary lowest-weight
			representations $\mathcal D(\Delta,s)$ of the conformal group with fixed dimension $\Delta$ and arbitrary
			integer spin $s$ contract to massless Poincaré particle representations. This requires
			combining the Inönü--Wigner contraction $\mathfrak{so}(2,3) \to \mathfrak{iso}(1,3)$ with an infinite boost, in which case massless scattering states naturally emerge in a carrollian basis. For $\Delta>s+1$, we find that $\mathcal D(\Delta,s)$
			contracts to the direct sum of massless particles of all spins $s'\leq s$. At the unitarity bound, lower-spin states become null so that
			$\mathcal D(s+1,s)$ contracts to a single massless particle representation carrying the two maximal helicities $J=\pm s$. In
			both cases, the inner product converges to the Poincaré-invariant one after a single
			helicity-independent rescaling, so that unitarity is preserved throughout the contraction.
		\end{abstract}
		
	\end{titlepage}
	
	{\hypersetup{linkcolor=black}
		\tableofcontents 
	}
	
	\section{Introduction}
	\label{section introduction}
	
	Since the early days of the AdS/CFT correspondence \cite{Aharony:1999ti}, it has been expected that a scattering process confined to a region much smaller than the AdS radius $\ell$ does not resolve the curvature, so that flat space S-matrix elements should be encoded in the boundary data of the dual conformal field theory \cite{Polchinski:1999ry,Susskind:1998vk,Giddings:1999qu,Giddings:1999jq}. A wide variety of prescriptions aiming to make precise this map from conformal correlators to scattering amplitudes have been put forward, using either Mellin, position, momentum, or carrollian variables \cite{Gary:2009ae,Heemskerk:2009pn,Fitzpatrick:2010zm,Penedones:2010ue,Fitzpatrick:2011jn,Fitzpatrick:2011ia,Fitzpatrick:2011hu,Fitzpatrick:2011dm,Paulos:2011ie,Paulos:2016fap,Hijano:2019qmi,Hijano:2020szl,Komatsu:2020sag,Li:2021snj,Gadde:2022ghy,vanRees:2022zmr,vanRees:2023fcf,Marotta:2024sce,Bagchi:2023fbj,Alday:2024yyj,Lipstein:2025jfj,Duary:2026hpb}. These prescriptions are mostly statements about observables,
	and what has remained largely unexplored is the contraction of the  Hilbert space itself: how unitary representations $\mathcal D(\Delta,s)$ of the conformal group, together with their inner product, reorganize into the Poincaré representations that define scattering states.
	In the first paper of this series \cite{Nguyen:2026rwt}, we initiated a systematic study of this flat limit directly at the level of the Hilbert space. The starting point is that, at the level of the symmetry algebra, the flat limit of AdS$_4$ corresponds to the Inönü–Wigner contraction $\mathfrak{so}(2,3)\to\mathfrak{iso}(1,3)$ \cite{Inonu:1953sp,Saletan:1961iqo,Mickelsson:1972fh}. The problem then consists in understanding how the unitary irreducible representations of the Poincaré group, that describe the scattering states entering the definition of a flat space S-matrix, emerge from the unitary lowest-weight representations $\mathcal D(\Delta,s)$ of the conformal group. 
	
	For massive particles, Evans established that massive Poincaré representations arise from a sequence of representations $\mathcal D(\Delta,s)$ with $\Delta\to\infty$ at fixed $m=\Delta/\ell$ \cite{Evans:1967zgs}. Fronsdal further observed that $\mathcal D(\Delta,s)$ admits a basis of `pseudo-momentum' states that reduce to Wigner's momentum eigenstates under the contraction, which he described in detail for scalar and spin-$\frac12$ representations \cite{Fronsdal:1965zzb,Fronsdal:1974ew,Fronsdal:1975eq}. In \cite{Nguyen:2026rwt}, we extended this construction to representations $\mathcal D(\Delta,s)$ of arbitrary dimension $\Delta$ and integer spin $s$. We also identified a simple conformal map between radially quantized conformal fields in $\mathbb R^3$ and pseudo-momentum states on the mass hyperboloid $\mathbb{H}^3$, such that the interior and exterior of the unit ball correspond to ingoing and outgoing particles, respectively. This Hilbert-space description parallels the recent analysis of bulk wavefunctions in \cite{Berenstein:2025tts,Berenstein:2025qhb}.
	
	The situation is markedly different for massless particles. Fronsdal observed long ago that, under the contraction $\mathfrak{so}(2,3)\to\mathfrak{iso}(1,3)$, any unitary representation $\mathcal D(\Delta,s)$ at fixed $\Delta$ yields a massless representation in the flat limit \cite{Fronsdal:1974ew,Fronsdal:1975eq}. This observation directly follows from the standard relation 
	\begin{equation}
		\label{ads dictionary}
		m^2 = (\Delta + s - 2)(\Delta - s - 1)/\ell^{2} \; \stackrel{\ell \to \infty}{\longrightarrow} \; 0\,, \qquad (\text{fixed }\Delta, s\geq 1)\,,
	\end{equation}
	which determines the bulk field mass in terms of the scaling dimension $\Delta$ and integer spin $s\geq 1$ of the dual conformal multiplet. Because of this universal mass vanishing at fixed $\Delta$ in the flat limit, it was initially not clear which representations deserved to be called a `massless particle in AdS'. Indeed, the various characterizations of masslessness in flat spacetime, such as gauge invariance, conformal invariance and null-cone propagation, no longer coincide in AdS \cite{Deser:1983tm,Deser:1983mm}. Historically, this was resolved by singling out the short multiplet representations $\mathcal D(s+1,s)$ at the unitarity bound, which are distinguished by gauge invariance of the corresponding AdS field equations \cite{Fronsdal:1975eq,Fronsdal:1978vb} and by their lift to unitary representations of the higher-dimensional conformal isometry group of the AdS bulk spacetime \cite{Angelopoulos:1980wg,Binegar:1981xc}.\footnote{For $s=0$, the corresponding role is played by the direct sum
		$\mathcal D(1,0)\oplus\mathcal D(2,0)$, which also extends to a unitary representation of the
		conformal isometry group of AdS$_4$, and which was contracted to a massless scalar in \cite{Angelopoulos:1980wg,Binegar:1981xc}. We will not deal with this particular case here.} In that context, a contraction of $\mathcal D(s+1,s)$ to a unitary Poincaré representation was constructed by exploiting this conformal extension, yielding a massless particle of a single helicity, either $J=+s$ or $J=-s$ \cite{Angelopoulos:1980wg}. Both helicities could only be obtained within a non-integrable representation of the Poincaré algebra, in which the energy operator fails to be self-adjoint, and it was conjectured that no contraction of the type considered there yields a unitary massless particle carrying both helicities \cite{Angelopoulos:1980wg}. Because it fails to reproduce massless particles with both helicities, their contraction does not account for the mechanism by which a gauge field in AdS reduces to a gauge field in flat spacetime. As we will see, this obstruction is specific to the framework of \cite{Angelopoulos:1980wg} and is lifted once the contraction is formulated in terms of infinitely boosted states with a rescaled inner product.
	
	In this paper, we implement the Inönü--Wigner contraction at the level of the representations, in such a way that unitary conformal representations $\mathcal D(\Delta,s)$ contract to unitary Poincaré representations describing \textit{massless} particles. For generic $\Delta$ above the unitarity bound, we obtain
	\begin{equation}
		\label{result generic}
		\mathcal{D}(\Delta,s) \qquad \stackrel{\ell \to \infty}{\longrightarrow}\qquad \bigoplus_{s'=0}^s\, \mathcal U_{s'}\,, \qquad (\Delta > s+1)\,,
	\end{equation}
	where $\mathcal{U}_{s'}$ denotes the Poincaré UIR describing a massless particle of spin $s'$, with both helicities $J=\pm s'$ included for $s'\geq 1$ so that it is also invariant under parity. At the unitarity bound $\Delta=s+1$, the same contraction procedure instead yields
	\begin{equation}
		\label{result short}
		\mathcal{D}(s+1,s) \qquad \stackrel{\ell \to \infty}{\longrightarrow}\qquad  \mathcal U_{s}\,,
	\end{equation}
	in agreement with the expectation that spin-$s$ gauge fields in AdS reduce to spin-$s$ gauge
	fields in Minkowski spacetime. Both results realize the branching rule
	$\operatorname{SO}(3)\downarrow\operatorname{SO}(2)$ of the lowest-weight multiplet onto the massless little group, with the lower spins $s'<s$ projected out for short multiplets. The branching underlying \eqref{result short} was first identified by Brink, Metsaev and Vasiliev in their analysis of the flat limit of massless fields in AdS$_{d+1}$ \cite{Brink:2000ag}, and was later formulated in terms of characters in \cite{Basile:2016aen}. The contraction described here realizes \eqref{result short} directly at the level of the Hilbert space for $d=3$, and extends it to \eqref{result generic} above the unitarity bound, where no component decouples and the branching is complete. Although we explicitly restrict our analysis to $d=3$, we believe that it can be generalized to arbitrary $d$ and that the results follow the branching rules $\operatorname{SO}(d)\downarrow\operatorname{SO}(d-1)$ identified in \cite{Brink:2000ag}.
	
	Two key elements underlie the contractions \eqref{result generic}--\eqref{result short}. The first is the pseudo-momentum basis of $\mathcal{D}(\Delta,s)$ introduced by Fronsdal \cite{Fronsdal:1965zzb,Fronsdal:1974ew}, which we generalized to arbitrary spin in \cite{Nguyen:2026rwt}. The second is the observation by Berenstein and Simon that obtaining non-vanishing momenta at fixed $\Delta$ in the flat limit requires implementing an infinite boost \textit{simultaneously} \cite{Berenstein:2025tts}.\footnote{Combining this observation with the infinite boost contraction of massive particles established in \cite{Nguyen:2026yho}, we were able to anticipate the particle content of \eqref{result generic} in \cite{Nguyen:2026rwt}.} This double scaling limit was analyzed at the level of bulk wavefunctions in \cite{Berenstein:2025tts,Berenstein:2025qhb}. It was found to produce massless wavefunctions that are not momentum eigenstates, together with longitudinal polarizations displaying divergent behavior. Working directly at the level of the Hilbert space rather than in terms of bulk field wavefunctions, we will not encounter such problematic behavior. The divergence resides in the polarization tensors that intertwine covariant fields with particle states, rather than in the states themselves. This distinction already appears in the infinite boost limit of massive Poincaré particles \cite{Nguyen:2026yho}. In that sense, the result \eqref{result generic} can be understood as the AdS counterpart of the Stückelberg formulation of massive fields in the high energy limit; see \cite{Nguyen:2026yho}.
	
	The paper is organized as follows. In Section~\ref{section 2}, we review the description of the unitary representations $\mathcal D(\Delta,s)$ in terms of Verma modules and radially quantized conformal fields in $\mathbb R^3$, and in terms of the pseudo-momentum basis in $\mathbb{H}^3$, as well as the conformal map relating the latter. In Section~\ref{section 3}, we implement the flat and infinite boost limits simultaneously, first in the pseudo-momentum basis and then in radial quantization. We show that the resulting states are not momentum eigenstates, but are rather \textit{carrollian} states familiar from Carrollian Holography \cite{Nguyen:2025zhg}. We derive the action of the Poincaré generators on the resulting carrollian states, decompose them into helicity components, and discuss the shortening condition at the unitarity bound. In Section~\ref{section inner product}, we show that the inner product on $\mathcal D(\Delta,s)$ converges to the Poincaré-invariant inner product, first for zero spin and then for integer spin, and we identify the null states of the short multiplets. We conclude in Section~\ref{section discussion} with a comparison against the existing literature, and some future directions. Details of the proof of the spin decomposition are given in Appendix~\ref{appendix general spin}. 
	
	\section{Unitary representations of the conformal group}
	\label{section 2}
	We first collect some elements of the representation theory of the conformal group
	$\operatorname{SO}(2,3)$ that are used in the rest of this work, referring to \cite{Nguyen:2026rwt} and references therein for further details. The conformal algebra $\mathfrak{so}(2,3)$ is given as
	\begin{equation}
		\left[L_{AB},L_{CD}\right]=i\left(\eta_{AC} L_{BD}+\eta_{BD}L_{AC}-\eta_{AD} L_{BC}-\eta_{BC} L_{AD} \right)\,,
	\end{equation}
	with $A=0,i,4$ and $i=1,2,3$, and where the nonzero components of the flat metric tensor $\eta_{AB}$ are given by
	\begin{equation}
		\eta_{00}=\eta_{44}=-1\,, \qquad \eta_{ij}=\delta_{ij}\,.
	\end{equation}
	Within a unitary representation, the generators are hermitian in the chosen conventions,
	\begin{equation}
		\label{hermiticity condition}
		L_{AB}^\dagger=L_{AB}\,.
	\end{equation}
	
	\subsection{Verma modules}
	Unitary irreducible representations of $\operatorname{SO}(2,3)$ are built from finite-dimensional irreducible representations of its maximal
	compact subgroup $\operatorname{SO}(2)\times\operatorname{SO}(3)$, generated by $D\equiv L_{40}$ and
	$L_{ij}$. The lowest-weight multiplet $|\Delta\rangle_I$ spanning the carrier space of this finite-dimensional representation satisfies
	\begin{equation}
		\label{primary states}
		D|\Delta\rangle_I = \Delta\,|\Delta\rangle_I\,, \qquad
		L_{ij}|\Delta\rangle_I = [L_{ij}]^{(s)}|\Delta\rangle_I\,,
	\end{equation}
	i.e., it has definite scaling weight $\Delta$ and carries a spin-$s$ representation of the rotation group. Here, we restrict to integer spin $s$, and encode spin-$s$ multiplets into totally
	symmetric traceless tensors $T_I=T_{i_1\dots\, i_s}$ of rank $s$, with multi-index $I = i_1\dots i_s$. The rotation generators $L_{ij}$ act on rank-$s$ tensors through the Leibniz rule,
	\begin{equation}
		\label{Leibniz rule}
		[L_{ij}]^{(s)} T_{k_1\dots\, k_s}= \sum_{r=1}^s [L_{ij}]\indices{_{k_r}^{l_r}}\, T_{k_1\dots\, l_r\, \dots\, k_s}\,,
	\end{equation}
	where $l_r$ stands in the $r$-th slot, and $[L_{ij}]$ is the vector representation 
	\begin{equation}
		[L_{ij}]\indices{_k^l} = i\,(\delta_{ik}\delta\indices{_j^l} - \delta_{jk}\delta\indices{_i^l})\,.
	\end{equation}
	These states are normalized as ${}_I\langle\Delta|\Delta\rangle_J = \delta_{IJ}$, where $\delta_{IJ}$ denotes the identity on totally symmetric traceless tensors. The remaining generators are
	arranged into raising and lowering operators conjugate to each other,
	\begin{equation}
		\label{ladder operators}
		L_i^\pm \equiv L_{i4} \mp i L_{i0}\,, \qquad \qquad (L^\pm_i)^\dagger = L^\mp_i\,.
	\end{equation}
	The primary states are annihilated by the lowering operators,
	\begin{equation}
		\label{primary condition}
		L^-_i |\Delta\rangle_I =0\,,
	\end{equation}
	and the parabolic Verma module $\mathcal V_{\Delta,s}$ is spanned by the descendants $L^+_{i_1}\dots L^+_{i_n}|\Delta\rangle_I$ for all values of $n \in \mathbb{N}$.
	
	The Verma module $\mathcal V_{\Delta,s}$ is unitary and irreducible provided
	\begin{equation}
		\label{unitarity bounds}
		\Delta > s+1 \quad (s\geq1)\,, \qquad \Delta > \tfrac12 \quad (s=0)\,.
	\end{equation}
	For $s\geq1$, the bound follows from the level-one states
	$\sum_i L^+_i|\Delta\rangle_{i i_2\dots i_s}$ whose norm is proportional to
	$(\Delta-s-1)$. At $\Delta=s+1$, this norm is zero and these states, which generate a submodule $\mathcal V_{s+2,s-1}\subset\mathcal V_{s+1,s}$, must be quotiented out. The unitary irreducible representation at $\Delta=s+1$ is the quotient
	\begin{equation}
		\mathcal D(s+1,s)=\mathcal V_{s+1,s}/\mathcal V_{s+2,s-1}\,,
	\end{equation}
	defined by the shortening condition
	\begin{equation}
		\label{null state}
		\sum_i L_i^+ |\Delta\rangle_{ii_2\dots i_s}=0\,, \qquad (\Delta=s+1)\,.
	\end{equation}
	For $s=0$, the module saturating the bound is also reducible, and its irreducible quotient is
	Dirac's singleton \cite{Dirac:1963ta,Flato:1978qz}.
	
	\subsection{Radial quantization}
	\label{sec 2.2}
	States of $\mathcal V_{\Delta,s}$ are in one-to-one correspondence with local tensor operators on
	the Euclidean space $\mathbb R^3$,\footnote{The correspondence is with Euclidean rather than
		Lorentzian tensor fields, since the rotation group $\operatorname{SO}(3)\subset\operatorname{SO}(2)\times\operatorname{SO}(3)$
		must act unitarily.}
	\begin{equation}
		\label{local conformal fields}
		|\Delta \rangle_{I} =O^\Delta_{I}(0)|0\rangle\,, \qquad
		O^\Delta_I(\vec x)=e^{x^i L_i^+}\, O^\Delta_I(0)\, e^{-x^i L_i^+}\,, \qquad |\vec x|<1\,,
	\end{equation}
	where the restriction to the unit ball ensures normalizability of the corresponding states \cite{Nguyen:2026rwt}. The conformal generators act
	as
	\begin{equation}
		\label{generator action on fields}
		\begin{split}
			\left[L_i^+, O^\Delta_I(\vec x)\right]&=\partial_i O^\Delta_I(\vec x)\,,\\
			\left[L_i^-, O^\Delta_I(\vec x)\right]&=\big(2\Delta x_i+2x_i x^j \partial_j-x^2 \partial_i-2i x^j[L_{ij}]^{(s)} \big)\,O^\Delta_I(\vec x)\,,\\
			\left[D, O^\Delta_I(\vec x)\right]&=(\Delta+x^i\partial_i )\,O^\Delta_I(\vec x)\,,\\
			\left[L_{ij}, O^\Delta_I(\vec x)\right]&=\big([L_{ij}]^{(s)}+i(x_i \partial_j-x_j\partial_i)\big)\,O^\Delta_I(\vec x)\,.
		\end{split}
	\end{equation}
	In this language, the shortening condition \eqref{null state} becomes the current conservation equation
	\begin{equation}
		\label{conservation equation}
		\partial^i O_{ii_2\dots i_s}(\vec x)=0\,, \qquad (\Delta=s+1)\,,
	\end{equation}
	which holds throughout the unit ball.
	
	Hermiticity \eqref{hermiticity condition} is implemented in $\mathbb R^3$ by the inversion
	$\mathcal{R}:\vec x\mapsto\vec x/x^2$ with $x\equiv |\vec x|$, 
	\begin{equation}
		\mathcal{R} L_i^\pm \mathcal{R}=L_i^\mp\,.
	\end{equation}
	This implies that the operator conjugate to $O_I(\vec x)$ is 
	\begin{equation}
		\label{conjugate operator}
		[O_I(\vec x)]^\dagger=\mathcal{R} O_I(\vec x) \mathcal{R}=x^{-2\Delta}\, \mathcal{I}\indices{_I^J}(\vec x)\, O_J(\vec x/x^2)\,,
	\end{equation} 
	where $\mathcal I_{IJ}(\vec x)$ is
	the inversion tensor, defined as the totally symmetric traceless part of the $s$-fold product of
	\begin{equation}
		\label{inversion tensor}
		\mathcal{I}_{ij}(\vec x)=\delta_{ij}-\frac{2 x_i x_j}{x^2}\,.
	\end{equation}
	It satisfies $\mathcal I_{IJ}(f\vec x)=\mathcal I_{IJ}(\vec x)$ for any non-vanishing $f$, and
	$\mathcal I\indices{_I^K}(\vec x)\,\mathcal I_{KJ}(\vec x)=\delta_{IJ}$. Since normalizable states (kets) correspond to operator insertions inside the unit ball, conjugate states (bras) correspond to operator insertions outside the unit ball.
	The overlap of two states is then given by 
	\begin{equation}
		\label{state overlap}
		\begin{split}
			\langle O_I(\vec x_1)|O_J(\vec x_2) \rangle&=\frac{\mathcal{I}\indices{_I^K}(\vec x_1)\, \mathcal{I}
				_{KJ}(\vec x_1/x_1^2-\vec x_2)}{(1-2\, \vec x_1 \cdot \vec x_2+x_1^2\, x_2^2)^\Delta}\,,
		\end{split}
	\end{equation}
	which is positive definite precisely when the unitarity bounds \eqref{unitarity bounds} are satisfied \cite{Osterwalder:1973dx,Osterwalder:1974tc,Kravchuk:2021kwe}.
	
	\subsection{Pseudo-momentum states}
	\label{sec 2.3}
	To describe the flat limit contraction of the lowest-weight representations, we use a
	third realization of $\mathcal V_{\Delta,s}$ in terms of \textit{pseudo-momentum states}, first introduced by Fronsdal for scalar representations \cite{Fronsdal:1965zzb,Fronsdal:1974ew}. Here, we review its recent generalization to arbitrary integer spin $s$ \cite{Nguyen:2026rwt}. 
	
	To this end, we introduce another basis of $\mathfrak{so}(2,3)$ generators, namely
	\begin{equation}
		\label{second generator basis}
		L_{\mu\nu}\,, \qquad \mathcal P_\mu\equiv c\, L_{\mu 4}\,, \qquad \mu,\nu=0,1,2,3\,,
	\end{equation}
	satisfying the commutation relations
	\begin{equation}
		\label{conformal algebra momentum basis}
		\begin{split}
			\left[L_{\mu\nu},L_{\rho\sigma}\right]&=i\left(\eta_{\mu \rho} L_{\nu \sigma}+\eta_{\nu\sigma}L_{\mu\rho}-\eta_{\mu\sigma} L_{\nu \rho}-\eta_{\nu\rho} L_{\mu\sigma} \right)\,,\\
			\left[L_{\mu\nu},\mathcal P_\rho \right]&=i \left(\eta_{\mu\rho}\mathcal P_\nu-\eta_{\nu\rho}\mathcal P_\mu \right)\,,\\
			\left[\mathcal P_\mu,\mathcal P_\nu \right]&=-i c^2 L_{\mu\nu}\,.
		\end{split}
	\end{equation}
	The limit $c\to 0$ describes the In\"on\"u--Wigner contraction
	$\mathfrak{so}(2,3)\to\mathfrak{iso}(1,3)$, where $\mathcal P_\mu$ become Poincar\'e translations. The
	parameter $c$ can be identified with the inverse AdS curvature radius $\ell^{-1}$ in \eqref{ads dictionary}, measured in some appropriate unit system (such as string units $\ell_s=1$).
	
	Since $\mathcal P_0 = -cD$, the primary-state conditions \eqref{primary states} are similar to those defining spin-$s$ massive particles in their rest frame, with rest energy 
	\begin{equation}
		\label{m=cD}
		m = c\,\Delta\,.
	\end{equation}
	Following Wigner's methodology, one can boost these primary states so as to cover the massive orbit $p^\mu p_\mu=-m^2$ \cite{Wigner:1939cj}. Let $L(\hat p)$ denote the standard boost taking the rest momentum $k^\mu=m\, \delta^\mu_0$ to the generic momentum $p^\mu=m\, \hat p^\mu$ with $\hat p^2=-1$. The pseudo-momentum states
	are then defined as
	\begin{equation}
		\label{pseudo momentum state}
		|p\rangle_I \equiv U(L(\hat p))\, |\Delta\rangle_I\,,
	\end{equation}
	in complete analogy with the standard Poincaré momentum states.
	The action of the Lorentz subgroup generated by $L_{\mu\nu}$ is exactly that of massive Poincar\'e particles, namely
	\begin{equation}
		\label{Lorentz transfo}
		U(\Lambda) |p\rangle_I = [W(\Lambda,p)]^{(s)} |\Lambda p\rangle_I\,, \qquad
		W(\Lambda,p)\equiv L^{-1}(\Lambda \hat p)\, \Lambda\, L(\hat p)\,,
	\end{equation}
	with $W(\Lambda,p) \in \operatorname{SO}(3)$ a Wigner rotation.
	Infinitesimally, this reads
	\begin{equation}
		\label{Lorentz generator pseudo-momentum action}
		\begin{split}
			L_{ij} |p\rangle_I&=\left[ [L_{ij}]^{(s)}+i \left(\hat p_i \frac{\partial}{\partial \hat p^j}-\hat p_j \frac{\partial}{\partial \hat p^i}\right)\right]|p\rangle_I\,,\\
			L_{0i}|p\rangle_I&=\left[ \frac{\hat p^j}{\hat p^0+1}\, [L_{ij}]^{(s)}+i \left(\hat p_0 \frac{\partial}{\partial \hat p^i}-\hat p_i \frac{\partial}{\partial \hat p^0}\right)\right]|p\rangle_I\,.
		\end{split}
	\end{equation}
	The action of the pseudo-momentum operator $\mathcal P_\mu$ is not diagonal however,
	\begin{equation}
		\label{P action}
		\mathcal P_\mu\, |p\rangle_I=\left[p_\mu+i c\, \hat p^\nu L_{\nu \mu} \right] |p\rangle_I\,.
	\end{equation}
	The pseudo-momentum states $|p\rangle_I$ therefore become momentum eigenstates only in the limit $c\to0$ at fixed
	$p_\mu$, reflecting the vanishing of $[\mathcal P_\mu,\mathcal P_\nu]=-ic^2L_{\mu\nu}$ in that limit. Their inner product is given by
	\begin{equation}
		\label{pseudo-momentum kernel}
		{}_I\langle p_1 | p_2 \rangle_J =  \left(\frac{2}{1-\hat p_1 \cdot \hat p_2} \right)^{\Delta}\, [W_{12}]_{IJ}\,,
		\qquad W_{12}\equiv W\big(L^{-1}(\hat p_1),p_2\big)\,,
	\end{equation}
	where $W_{12}$ is a rotation about the axis $\vec p_1\times\vec p_2$ \cite{Nguyen:2026rwt}. 
	
	In Part I of this series \cite{Nguyen:2026rwt}, we demonstrated that the pseudo-momentum states $|p\rangle_I$ become the standard momentum eigenstates of a massive Poincaré irrep of mass $m$ and spin $s$, provided we take the Inönü--Wigner contraction $c \to 0$ at fixed $m=c \Delta$. In other words, one has to take the scaling dimension $\Delta$ to infinity in the process. This way of implementing the contraction to massive Poincaré irreps is in agreement with Evans' old result \cite{Evans:1967zgs}. The purpose of the present work is to adapt this approach so as to obtain \textit{massless} Poincaré particles as a result of the flat limit contraction $c \to 0$.
	
	\subsection{From position to pseudo-momentum space}
\begin{figure}[H]
		\centering
		\begin{tikzpicture}[scale=1.5,>={Stealth}]
			\fill[blue!7] (-1.65,-1.55) rectangle (1.65,1.55);
			\node[anchor=north west] at (-1.65,1.55) {\footnotesize $\mathbb{R}^3$};
			\fill[orange!18] (0,0) circle (1);
			\fill[red!55,opacity=0.55,even odd rule] (0,0) circle (1) (0,0) circle (0.9);
			\fill[blue!55,opacity=0.55,even odd rule] (0,0) circle (1.1) (0,0) circle (1);
			\draw[thick,orange!50!red] (0,0) circle (1);
			\draw[orange!50!red,dashed] (-1,0) arc (180:360:1 and 0.25);
			\draw[orange!50!red,densely dotted] (1,0) arc (0:180:1 and 0.25);
			\node at (0,0.45) {\scriptsize $x<1$};
			\node at (1,-1.3) {\scriptsize $x>1$};
			\begin{scope}[shift={(5.6,0)}]
				\draw[gray,dashed] (-1.45,-1.45) -- (1.45,1.45);
				\draw[gray,dashed] (-1.45,1.45) -- (1.45,-1.45);
				\draw[->,gray] (0,-1.7) -- (0,1.8) node[above] {\scriptsize $\ph^0$};
				\fill[orange!18] plot[domain=-1.2:1.2,samples=40] (\x,{sqrt(0.25+\x*\x)}) -- cycle;
				\draw[thick,orange!50!red] plot[domain=-1.2:1.2,samples=40] (\x,{sqrt(0.25+\x*\x)});
				\fill[orange!10] (0,1.3) ellipse (1.2 and 0.2);
				\fill[red!55,opacity=0.55]
				(-0.923,1.05) arc (180:360:0.923 and 0.154)
				-- plot[domain=0.923:1.2,samples=10] (\x,{sqrt(0.25+\x*\x)})
				arc (0:-180:1.2 and 0.2)
				-- plot[domain=-1.2:-0.923,samples=10] (\x,{sqrt(0.25+\x*\x)}) -- cycle;
				\draw[thick,orange!50!red] (0,1.3) ellipse (1.2 and 0.2);
				\fill[blue!12] plot[domain=-1.2:1.2,samples=40] (\x,{-sqrt(0.25+\x*\x)}) -- cycle;
				\draw[thick,blue!60!black] plot[domain=-1.2:1.2,samples=40] (\x,{-sqrt(0.25+\x*\x)});
				\fill[blue!55,opacity=0.55]
				(-0.923,-1.05) arc (180:360:0.923 and 0.154)
				-- plot[domain=0.923:1.2,samples=10] (\x,{-sqrt(0.25+\x*\x)})
				arc (0:-180:1.2 and 0.2)
				-- plot[domain=-1.2:-0.923,samples=10] (\x,{-sqrt(0.25+\x*\x)}) -- cycle;
				\draw[thick,blue!60!black] (0,-1.3) ellipse (1.2 and 0.2);
				\node[right] at (1.4,1.3) {\scriptsize $\ph^0\to+\infty$};
				\node[right] at (1.4,-1.3) {\scriptsize $\ph^0\to-\infty$};
			\end{scope}
			\draw[->,thick,orange!50!red] (0.6,0.75) to[bend right=25] node[above,pos=0.5] {\scriptsize $x\to 1^-$} (4.45,1.15);
			\draw[->,thick,blue!60!black] (0.75,-0.75) to[bend left=25] node[below,pos=0.5] {\scriptsize $x\to 1^+$} (4.45,-1.15);
		\end{tikzpicture}
		\caption{The conformal map \eqref{momentum parametrisation} between $\mathbb R^3$ and the
			two-sheeted mass hyperboloid $\mathbb{H}^3$. The interior of the unit ball, where operator insertions create ket states, maps to
			future-directed pseudo-momenta, while the exterior, where operator insertions create bra states, maps to
			past-directed ones. The infinite boost limit described in Section~\ref{section 3}
			probes a thin shell enclosing the unit sphere, which is mapped to the
			asymptotia $\p^0\to \pm \infty$ of $\mathbb{H}^3$.}
		\label{fig:ball to hyperboloid}
	\end{figure}
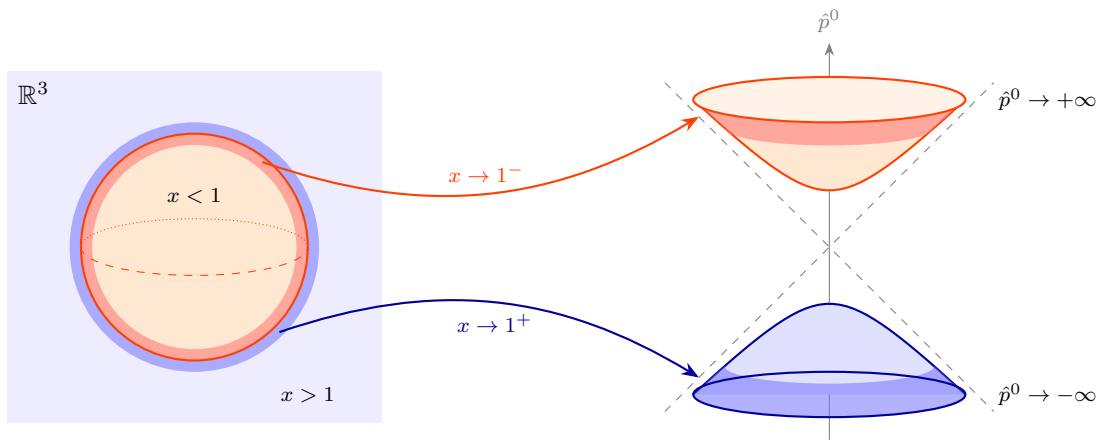

	There exists a strikingly simple relation between the pseudo-momentum states of Section~\ref{sec 2.3} and the Euclidean conformal fields of Section~\ref{sec 2.2}, given in terms of the conformal map between $\mathbb{R}^3$ and the momentum mass-shell $\mathbb{H}^3$ \cite{Nguyen:2026rwt}. More precisely, we have
	\begin{equation}
		\label{Psi(p) from O(x)}
		|p\rangle_I = (1-x^2)^{\Delta}\, O^\Delta_I(\vec x)|0\rangle\,, 
	\end{equation}
	where the change of coordinates is given by
	\begin{equation}
		\label{momentum parametrisation}
		\hat p^\mu(\vec x)=\left(\frac{1+x^2}{1-x^2}\,, \frac{2 \vec x}{1-x^2} \right)\,.
	\end{equation}
	The interior of the unit ball covers future-directed momenta, $\hat p^0>0$, while the exterior covers past-directed ones, and inversion acts as $\mathcal{R}:\hat p^\mu\mapsto-\hat p^\mu$. Within a correlation function, operator insertions inside the unit ball create the ket states, i.e., the incoming particles with their physical momenta, while insertions outside the unit ball create the bra states, i.e., the outgoing particles labeled by minus their physical momenta; see Figure~\ref{fig:ball to hyperboloid}. This matches the \textit{all-incoming} convention for scattering amplitudes.
	
	The identification \eqref{Psi(p) from O(x)} allows one to recover the overlap
	\eqref{pseudo-momentum kernel} from the inner product \eqref{state overlap}. On the one hand,
	\begin{equation}
		\left( \frac{(1-x_1^2)(1-x_2^2)}{1-2 \vec x_1 \cdot \vec x_2+x_1^2\, x_2^2}\right)^\Delta
		=\left( \frac{2}{1-\hat p_1 \cdot \hat p_2}\right)^\Delta\,,
	\end{equation}
	as directly follows from \eqref{momentum parametrisation}. On the other hand, the inversion tensors reproduce the
	Wigner rotation,
	\begin{equation}
		\label{II=W main}
		\mathcal{I}\indices{_I^K}(\vec x_1)\, \mathcal{I}_{KJ}(\vec x_1/x_1^2-\vec x_2)=[W_{12}]_{IJ}\,,
	\end{equation}
	as proven in Appendix B of \cite{Nguyen:2026rwt}. 
	
	\section{Massless particles in the flat limit}
	\label{section 3}
	In the flat limit contraction that yields massive Poincaré particles  \cite{Nguyen:2026rwt}, the limit $c \to 0$ is taken at fixed mass $m=c\, \Delta$, so that $\Delta\to\infty$ and the pseudo-momentum states become momentum
	eigenstates. Instead, here we wish to keep $\Delta$ fixed, so that $m\to0$ and massless Poincaré particles are obtained. However, retaining finite energy in the limit requires implementing at the same time an infinite boost to the pseudo-momentum states, as already suggested in \cite{Berenstein:2025qhb,Berenstein:2025tts} from the perspective of bulk wavefunctions. Perhaps surprisingly, we will see that the resulting massless states are \textit{not} momentum eigenstates. Rather, they correspond to massless Poincaré particles in a \textit{carrollian} basis.
	
	\subsection{Pseudo-momentum perspective}
	\label{section 3.1}
	To demonstrate this, we first parametrize the unit pseudo-momentum hyperboloid as
	\begin{equation}
		\label{momentum rapidity param}
		\hat p^\mu=(\cosh \eta, \sinh \eta\, \vec n)\,, \qquad \vec n^2=1\,,
	\end{equation}
	where $\eta \geq 0$ is the rapidity and $\vec n$ is a vector on the unit sphere $S^2$. Comparing with \eqref{momentum parametrisation}, this corresponds to a change of coordinate $x=|\vec x|=\tanh(\eta/2)$. We also introduce tangential derivatives to the sphere,
	\begin{equation}
		D_i\equiv (\delta_{ij}-n_in_j)\, \frac{\partial}{\partial n^j}\,, \qquad n^i D_i=0\,,
	\end{equation}
	where $\delta_{ij}-n_i n_j$ projects onto the tangent bundle of the sphere.
	In these coordinates, the action \eqref{Lorentz generator pseudo-momentum action} of the Lorentz subgroup  reads
	\begin{equation}
		\label{Lorentz action rapidity}
		\begin{split}
			L_{ij} |p\rangle_I &=\left[[L_{ij}]^{(s)}+i\left(n_i D_j-n_j D_i\right) \right] |p\rangle_I\,,\\
			L_{0i} |p\rangle_I &=\left[\tanh \left(\tfrac{\eta}{2}\right) n^j [L_{ij}]^{(s)}-i\,(n_i\, \partial_\eta+\coth \eta\, D_i) \right]|p \rangle_I\,,
		\end{split}
	\end{equation}
	while the action \eqref{P action} of the pseudo-momenta becomes
	\begin{equation}
		\label{P action rapidity}
		\begin{split}
			\mathcal P_0\, |p\rangle_I&=-c\big[\Delta \cosh \eta+ \sinh \eta\, \partial_\eta\big]|p\rangle_I\,,\\
			\mathcal P_i\, |p\rangle_I&=c\, \Big[\Delta \sinh \eta\, n_i+\cosh \eta\, n_i\, \partial_\eta+\frac{D_i}{\sinh \eta}-i \tanh \left(\tfrac{\eta}{2}\right) n^j [L_{ij}]^{(s)} \Big] |p\rangle_I\,.
		\end{split}
	\end{equation}
	It is clear that taking the flat limit contraction $c \to 0$ while keeping all other parameters fixed would yield a trivial action of the momenta $\mathcal P_\mu =0$, which is inappropriate for describing massless Poincaré particles. The way out, as suggested by Berenstein and Simon \cite{Berenstein:2025tts}, is to implement at the same time an infinite boost $\eta \to \infty$ so as to keep the momenta finite. For $e^\eta\gg1$, \eqref{Lorentz action rapidity}-\eqref{P action rapidity} yield
	\begin{equation}
		\label{P large eta expansion}
		\begin{split}
			\mathcal P_0\, |p\rangle_I&=-\frac{c\, e^\eta}{2}\left[\Delta +  \partial_\eta +O(e^{-2\eta})\right]|p\rangle_I \,,\\
			\mathcal P_i\, |p\rangle_I&=\frac{c\, e^\eta}{2}\left[n_i (\Delta  + \partial_\eta)+O(e^{-\eta}) \right] |p\rangle_I\,,\\
			L_{0i} |p \rangle_I &=\left[n^j [L_{ij}]^{(s)}-i\,(n_i\, \partial_\eta+ D_i)+O(e^{-\eta}) \right]|p \rangle_I\,,
		\end{split}
	\end{equation}
	while $L_{ij}|p\rangle_I$ is unchanged. 
	
	The action of the pseudo-momenta $\mathcal P_\mu$ can be made nonzero and finite in the limit $c\to0$ at fixed
	$\Delta$, provided we also keep fixed the rescaled coordinate $\tau$, defined as
	\begin{equation}
		\label{tau definition}
		\frac{1}{\tau}\equiv \frac{c\, e^\eta}{2}\,. \qquad \text{(fixed in the limit)}
	\end{equation}
	This amounts to implementing an infinite boost $\eta \to \infty$ with $e^{-\eta} \sim c$. 
	After the flat limit contraction, $\mathcal P_\mu$ become genuine Poincaré momenta, and using $\partial_\eta=-\tau \partial_\tau$, their action \eqref{P large eta expansion} becomes 
	\begin{equation}
		\label{Q on pseudo momentum states}
		\mathcal P_\mu\, |p\rangle_I 
		=  n_\mu \left(\Delta\, \tau^{-1}-\partial_\tau  \right) |p\rangle_I\,,
	\end{equation}
	where we introduced the null four-vector
	\begin{equation}
		n^\mu=(1,\vec n)\,, \qquad n^2=0\,.
	\end{equation}
	After the flat limit contraction, the pseudo-momentum states satisfy 
	\begin{equation}
		\mathcal{P}_\mu \mathcal{P}^\mu |p\rangle_I =0\,,
	\end{equation}
	which directly follows from $n^2=0$. Hence, the resulting Poincaré representation is \textit{massless with nontrivial momenta} as we intended.
	Unlike for the flat limit contraction to massive Poincaré particles \cite{Nguyen:2026rwt}, the states $|p\rangle_I$ do \textit{not} become momentum eigenstates. This is already visible in \eqref{P action}: at fixed $\Delta$, the coefficient
	$c\,\hat p^\nu = p^\nu/\Delta$ of the correction term remains finite in the limit $c \to 0$ at fixed $\tau$. However, the action of the momenta \eqref{Q on pseudo momentum states} does commute, $[\mathcal P_\mu,\mathcal P_\nu]=0$, consistently with the Inönü--Wigner contraction $\mathfrak{so}(2,3) \to \mathfrak{iso}(1,3)$ which we aimed to describe. For the Lorentz transformations, \eqref{P large eta expansion} reduces to
	\begin{equation}
		\label{Lorentz action flat limit}
		\begin{split}
			L_{ij} |p\rangle_I &=\left[[L_{ij}]^{(s)}+i\left(n_i D_j-n_j D_i\right) \right] |p\rangle_I\,,\\
			L_{0i} |p \rangle_I &=\left[n^j [L_{ij}]^{(s)}+i\,(n_i\, \tau \partial_\tau- D_i) \right]|p\rangle_I\,.
		\end{split}
	\end{equation}
	For the sake of clarity, it is best to label the states obtained from the flat limit contraction with the coordinates $(\tau, \vec n)$ that have been kept fixed, namely
	\begin{equation}
		\label{states tau definition}
		|\tau,\vec n\rangle_I \equiv (2\tau)^{-\Delta} \lim_{\substack{c \to 0\\ \text{fixed } (\tau,\vec n)}} |p \rangle_I\,.
	\end{equation}
	The rescaling by the factor $(2\tau)^{-\Delta}$ is introduced here for later convenience, and modifies \eqref{Q on pseudo momentum states} and \eqref{Lorentz action flat limit} accordingly so that
	\begin{equation}
		\label{Poincare action tau}
		\begin{split}
			\mathcal P_\mu\, |\tau,\vec n\rangle_I 
			&=  -n_\mu \partial_\tau  |\tau,\vec n\rangle_I\,,\\
			L_{ij} |\tau,\vec n\rangle_I &=\left[[L_{ij}]^{(s)}+i\left(n_i D_j-n_j D_i\right) \right] |\tau,\vec n\rangle_I\,,\\
			L_{0i} |\tau, \vec n \rangle_I &=\left[n^j [L_{ij}]^{(s)}+i\,(n_i\, (\tau \partial_\tau+\Delta)- D_i) \right]|\tau,\vec n\rangle_I\,.
		\end{split}
	\end{equation}
	
	\paragraph{Momentum and carrollian bases.}
	Next, we work out the relation between the states $|\tau,\vec n\rangle$ and the familiar massless momentum eigenstates $|q\rangle$, in the spinless case $s=0$. The treatment of spinning particles will be given in Section~\ref{section 3.3}.
	
	Let us denote by $|q\rangle$ the momentum eigenstates of a massless particle UIR of spin zero, where a generic massless momentum may be parametrized by
	\begin{equation}
		q^\mu(\omega,\vec n)=\omega\, n^\mu\,.
	\end{equation}
	We wish to work out the decomposition of $|\tau,\vec n\rangle$ onto this momentum basis. Since $\mathcal P_\mu$ in \eqref{Poincare action tau} leaves $\vec n$ invariant, we can look for a relation of the form
	\begin{equation}
		\label{transform generic}
		|\tau,\vec n\rangle= \int_0^\infty d\omega\, f(\omega,\tau)\, |q(\omega,\vec n)\rangle\,,
	\end{equation}
	with $f(\omega,\tau)$ some unknown intertwining kernel.
	Matching the action of $\mathcal P_\mu$ on both sides of \eqref{transform generic} using \eqref{Poincare action tau} yields the constraint
	\begin{equation}
		\omega f(\omega,\tau)=-\partial_\tau f(\omega,\tau)\,,
	\end{equation}
	solved by
	\begin{equation}
		\label{f solution 1}
		f(\omega,\tau)= e^{-\omega \tau} g(\omega)\,,
	\end{equation}
	where $g(\omega)$ is yet undetermined. This undetermined function is fixed by boosting along $\vec n$, under which
	$\eta\mapsto\eta+\eta_0$, such that
	$\tau \mapsto e^{-\eta_0}\tau$ and
	$\omega \mapsto e^{\eta_0} \omega$. Applying this to both sides of
	\eqref{transform generic} yields the constraint
	\begin{equation}
		f(\omega,\lambda\tau)=\lambda^{1-\Delta} f(\lambda\omega,\tau)\,, \qquad (\lambda=e^{-\eta_0})
	\end{equation}
	which determines
	\begin{equation}
		g(\omega)\propto \omega^{\Delta-1}\,.
	\end{equation}
	The change of basis \eqref{transform generic} is therefore given by
	\begin{equation}
		\label{carrollian states}
		|\tau,\vec n\rangle =\mathcal{N}\, \int_0^\infty d\omega\, \omega^{\Delta-1} e^{-\omega \tau}\, |q(\omega,\vec n)\rangle \,,
	\end{equation}
	up to some normalization constant $\mathcal{N}$. 
	In other words, $|\tau,\vec n\rangle$ is the \textit{Laplace transform} of $\omega^{\Delta-1} |q(\omega,\vec n)\rangle$ at fixed $\vec n$.
	Since $\omega\geq0$, the right-hand side is analytic for $\text{Re}\,\tau>0$, and the
	transform is inverted by the Bromwich integral
	\begin{equation}
		\label{inverse transform}
		|q(\omega,\vec n)\rangle = \frac{\omega^{1-\Delta}}{2\pi i\, \mathcal N} 
		\int_{\gamma-i\infty}^{\gamma+i\infty} d\tau\,e^{\omega\tau}\, |\tau,\vec n\rangle\,,
	\end{equation}
	for any real $\gamma>0$. For $\omega<0$,
	the contour can be closed in the half-plane $\text{Re}\,\tau>0$ and the integral vanishes, consistently with positivity of the energy.
	
	We are now only one step away from witnessing the emergence of the \textit{carrollian} basis of massless Poincaré particles \cite{Banerjee:2018gce,Nguyen:2023vfz}. Indeed, it is natural to perform a Wick rotation $\tau=-i u+0^+$ such that the integral \eqref{inverse transform} runs along the real axis, 
	\begin{equation}
		\label{inverse transform bis}
		|q(\omega,\vec n)\rangle = \frac{\omega^{1-\Delta}}{2\pi\, \mathcal N} 
		\int_{-\infty}^{\infty} du\,e^{-i\omega(u+i0^+)}|u,\vec n\rangle\,, 
	\end{equation}
	with $|u,\vec n\rangle\equiv |\tau,\vec n\rangle|_{\tau=-iu+0^+}$. In this case, \eqref{carrollian states} reduces to the \textit{modified Mellin transform} familiar from Carrollian Holography \cite{Banerjee:2018gce,Banerjee:2019prz,Bagchi:2022emh,Donnay:2022wvx,Nguyen:2023vfz},
	\begin{equation}
		\label{carrollian states bis}
		|u,\vec n\rangle=\mathcal{N}\,\int_0^\infty d\omega\, \omega^{\Delta-1} e^{i\omega (u+i0^+)}\, |q(\omega,\vec n)\rangle \,,
	\end{equation}
	i.e., the states $|u,\vec n\rangle$ are carrollian conformal states, with $(u,\vec n)$ playing the role of local coordinates at null infinity $\scri \simeq \mathbb{R}_u \times S^2$ (in the round sphere conformal frame).

	\subsection{Radial-quantization perspective}
	\label{section 3.2}
	An alternative route consists of doing the flat limit contraction directly from the flat Euclidean $\mathbb{R}^3$ where radial quantization is defined.  As illustrated in Figure~\ref{fig:ball to hyperboloid}, the infinite boost limit corresponds to the limit $x \to 1$ in $\mathbb{R}^3$. To implement it in practice, we introduce the radial foliation
	\begin{equation}
		\label{radial foliation}
		\vec x=r\, \vec n\,, \qquad |\vec n|=1\,.
	\end{equation}
	Using
	\begin{equation}
		\label{Jacobian derivatives}
		\partial_i = n_i\,\partial_r + \frac{1}{r}\,D_i\,, \qquad
		x^i\partial_i = r\,\partial_r\,, \qquad
		x_i\partial_j - x_j\partial_i = n_i D_j - n_j D_i\,,
	\end{equation}
	the action \eqref{generator action on fields} of the $\mathfrak{so}(2,3)$ generators reads, dropping the $\Delta$ label on $O^\Delta_I$ for convenience,
	\begin{equation}
		\begin{split}
			\big[L^+_i, O_I\big]&= \left(n_i\,\partial_r + r^{-1} D_i\right) O_I\,, \\
			\big[L^-_i,  O_I\big]&= \left(2\Delta r\, n_i + r^2 n_i\,\partial_r - r D_i- 2i r\, n^j [L_{ij}]^{(s)}\right)O_I\,, \\
			\big[D\,, O_I\big]&= \left(\Delta + r\partial_r\right)O_I\,, \\
			\big[L_{ij}\,, O_I\big]&= \left([L_{ij}]^{(s)} + i(n_i D_j - n_j D_i)\right)O_I\,,
		\end{split}
	\end{equation}
	or, in terms of the generators \eqref{second generator basis},
	\begin{equation}
		\label{P L action in radial coordinate}
		\begin{split}
			\big[\mathcal P_0\,, O_I\big]&= -c\left(\Delta + r\,\partial_r\right)O_I\,, \\
			\big[\mathcal P_i\,, O_I\big]&= \frac{c}{2}\left((1+r^2) n_i\,\partial_r +r^{-1}(1-r^2) D_i+ 2\Delta r\, n_i - 2i r\, n^j [L_{ij}]^{(s)}\right)O_I\,, \\
			\big[L_{0i}\,, O_I\big]&=\frac{1}{2}\left(2r n^j [L_{ij}]^{(s)} + 2i\Delta r\, n_i+ i(r^2-1) n_i\,\partial_r- ir^{-1}(r^2+1)D_i\right) O_I\,,\\
			\big[L_{ij}\,, O_I\big]&= \left([L_{ij}]^{(s)} + i(n_i D_j - n_j D_i)\right)O_I\,.
		\end{split}
	\end{equation}
	The Euclidean time $\tau$ defined in \eqref{tau definition} is related to $r$ through 
	\begin{equation}
		\label{r tau relation}
		r=\tanh \frac{\eta}{2}=\frac{2-c \tau}{2+c \tau}=1-c \tau+O(c^2)\,,
	\end{equation}
	such that \eqref{P L action in radial coordinate} becomes, in the $c \to 0$ limit,
	\begin{equation}
		\label{Poincare action tau operator}
		\begin{split}
			\big[\mathcal P_\mu\,, O_I\big]&=-n_\mu \partial_\tau\,  O_I\,, \\
			\big[L_{0i}\,, O_I\big]&=\left( n^j [L_{ij}]^{(s)} + i\left(n_i(\tau \partial_\tau+\Delta)-D_i \right)\right) O_I\,,\\
			\big[L_{ij}\,, O_I\big]&= \left([L_{ij}]^{(s)} + i(n_i D_j - n_j D_i)\right)O_I\,.
		\end{split}
	\end{equation}
	This is precisely equivalent to \eqref{Poincare action tau} which was obtained by performing the flat limit contraction in the pseudo-momentum state basis. More precisely, using \eqref{Psi(p) from O(x)} and the definition \eqref{states tau definition}, together with the change of coordinates \eqref{radial foliation} and \eqref{r tau relation}, we find
	\begin{equation}
		\label{3.29}
		|\tau,\vec n\rangle_I=\lim_{c \to 0} c^\Delta\, O_I(\tau,\vec n)|0\rangle\,.
	\end{equation}
	
	In the scalar case $s=0$, the Poincaré transformations \eqref{Poincare action tau operator} are those of a carrollian conformal field of scaling dimension $\Delta$ locally defined on a copy of null infinity $\scri \simeq \mathbb{R}_u \times S^2$, modulo the Wick rotation $\tau=-i u+0^+$ introduced earlier. The situation for nonzero spin $s \neq 0$ is less transparent, as the operators $O_I$ carry $\operatorname{SO}(3)$ spin indices $I=i_1\dots \, i_s$, while carrollian conformal fields naturally carry $\operatorname{SO}(2)$ spin indices or helicity labels, which are indeed those appropriate to describing spinning massless particles. Before untangling these spin degrees of freedom in Section~\ref{section 3.3}, we will first implement a conformal map from $S^2$ to $\mathbb{R}^2$ through stereographic projection. The resulting flat conformal frame is the one usually adopted to describe carrollian conformal fields \cite{Nguyen:2025zhg}.
	
	\paragraph{Flat conformal frame.} We introduce a stereographic projection of the unit sphere via
	\begin{equation}
		\vec n = \Big(\frac{2y^a}{1+y^2}\,,\ \frac{1-y^2}{1+y^2}\Big)\,, 
		\qquad y^2 \equiv \delta_{ab}\, y^a y^b\,,
	\end{equation}
	with indices $a,b$ running over $1,2$, and implement the conformal transformation to the flat conformal frame via
	\begin{equation}
		\label{flat conformal frame}
		\hat O_I(t,y)\equiv  \Omega^{-\Delta}\, O_I(\tau,\vec n)\,, 
	\end{equation}
	where
	\begin{equation}
		t \equiv \Omega\,\tau\,, \qquad \Omega \equiv \frac{1+y^2}{\sqrt 2}\,.
	\end{equation}
	In the new coordinate system, we recognize $\Omega\, n^\mu$ as the null vector familiar from Celestial and Carrollian Holography, namely
	\begin{equation}
		\hat q^\mu(y)\equiv \Omega\, n^\mu = \frac{1}{\sqrt 2} \left(1+y^2\,, 2y^a\,, 1-y^2\right)\,.
	\end{equation}
	Using the relations
	\begin{equation}
		n_i - D_i\ln\Omega = (y^1,y^2,1)\,, \qquad
		D_a y^b = \frac{1}{2}(1+y^2)\,\delta_a^b - y_a y^b\,, \qquad
		D_3 y^b = -y^b\,,
	\end{equation}
	we obtain the action of the $\mathfrak{iso}(1,3)$ generators on the fields of the flat conformal frame $\mathbb{R}_t \times \mathbb{R}^2$, namely
	\begin{equation}
		\begin{split}
			[\mathcal P_\mu\,,\hat O_I] &= -\hat q_\mu\, \partial_t \hat O_I\,,\\
			[L_{ab}\,,\hat O_I] &= \left([L_{ab}]^{(s)}+ i\,(y_a\partial_b - y_b\partial_a)\right)\hat O_I\,,\\
			[L_{3a}\,,\hat O_I] &= \left([L_{3a}]^{(s)}+ i y_a\,(t\partial_t + y^b\partial_b + \Delta)+ \tfrac{i}{2}(1-y^2)\partial_a\right)\hat O_I\,,\\
			[L_{0a}\,,\hat O_I] &= \left(n^j[L_{aj}]^{(s)}+ iy_a\,(t\partial_t + y^b\partial_b + \Delta)- \tfrac{i}{2}(1+y^2)\partial_a\right)\hat O_I\,,\\
			[L_{03}\,,\hat O_I] &= \left(n^j[L_{3j}]^{(s)}+ i\,(t\partial_t + y^b\partial_b + \Delta)\right)\hat O_I\,.
		\end{split}
	\end{equation}
	To make contact with the description of carrollian conformal fields given in \cite{Nguyen:2023vfz}, we re-express these in the carrollian basis of $\mathfrak{iso}(1,3)$ given by\footnote{Through this identification, it can be seen that the Inönü--Wigner contraction described in this work is manifestly the same as the ultra-relativistic/carrollian contraction described in Appendix~A of \cite{Nguyen:2023vfz}. This is the reason we denote by $c$ the contraction parameter, as it can be identified with the speed of light in a three-dimensional Minkowski spacetime on which $\operatorname{SO}(2,3)$ is realized as the group of conformal isometries.} 
	\begin{equation}
		\label{carrollian generators}
		H=\tfrac{1}{\sqrt{2}}(\mathcal P_0-\mathcal P_3)\,, \qquad K=\tfrac{1}{\sqrt{2}}(\mathcal P_0+\mathcal P_3)\,, \qquad B_a=-\tfrac{1}{\sqrt{2}} \mathcal P_a\,,
	\end{equation}
	for the abelian subalgebra, and\footnote{Although we use the same letter $D$ to denote them both, note that the carrollian dilation $D=L_{30}$ in \eqref{carrollian generators bis} is distinct from the $\mathfrak{so}(2,3)$ dilation operator that appeared earlier and which was defined by $D=L_{40}$.} 
	\begin{equation}
		\label{carrollian generators bis}
		L_{ab}\,, \qquad  P_a=L_{0a}-L_{3a}\,, \qquad K_a=L_{0a}+L_{3a}\,, \qquad D=L_{30}\,,
	\end{equation}
	for the Lorentz subalgebra, which yields
	\begin{equation}
		\label{carrollian action 1}
		\begin{split}
			[H, \hat O_I]&=\partial_t \hat O_I\,,\\
			[K,\hat O_I]&=y^2 \partial_t \hat O_I\,,\\
			[B_a,\hat O_I]&=y_a \partial_t \hat O_I\,,\\
			[D,\hat O_I]&=-\left(n^j[L_{3j}]^{(s)}+ i\,(t\partial_t + y^b\partial_b + \Delta)\right)\hat O_I\,,\\
			[P_a, \hat O_I]&=\left(n^j[L_{aj}]^{(s)}+[L_{a3}]^{(s)}- i\partial_a\right)\hat O_I\,,\\
			[K_a,\hat O_I]&=\left(n^j[L_{aj}]^{(s)}-[L_{a3}]^{(s)}+ 2iy_a\,(t\partial_t + y^b\partial_b + \Delta)- iy^2\partial_a\right)\hat O_I\,,\\
			[L_{ab}\,,\hat O_I] &= \left([L_{ab}]^{(s)}+ i\,(y_a\partial_b - y_b\partial_a)\right)\hat O_I\,.
		\end{split}
	\end{equation}
	Comparison with \cite{Nguyen:2023vfz} shows that the orbital part of the transformations indeed coincides with that of the carrollian conformal fields, provided we again introduce the Wick rotation $t=-iu+0^+$, where $u$ is now the retarded time in the flat conformal frame in which $\scri \simeq \mathbb{R}_u \times \mathbb{R}^2$. The spin contributions, however, still need to be sorted out. In particular, carrollian conformal fields $\hat O_{a_1...\,a_s}$ carry $\operatorname{SO}(2)$ indices encoding the corresponding  massless spin-$s$ degrees of freedom, while the fields $\hat O_I$ carry $\operatorname{SO}(3)$ tensor indices $I=i_1...\,i_s$ inherited from the lowest-weight multiplet structure.  
	
	\subsection{Helicity decomposition}
	\label{section 3.3}
	
	Let us first illustrate the decomposition of $\hat O_I$ into $\operatorname{SO}(2)$ tensors with the simplest nontrivial case of the vector representation ($s=1$), dropping the `hat' notation for convenience. We will then return to the case of generic integer spin $s$. 
	
	For $s=1$ the multi-index reduces to a regular index, with the representation of the rotation generators given by
	\begin{equation}
		[L_{ij}]^{(1)}O_k = i\,(\delta_{ik}O_j - \delta_{jk}O_i)\,.
	\end{equation}
	To untangle the $\operatorname{SO}(2)$ spin multiplets, we introduce an orthonormal frame adapted to the flat conformal frame, given by the triad $(\vec e_1,\vec e_2,\vec n)$, with
	\begin{equation}
		\vec e_a=\frac{1+y^2}{2}\,\partial_a \vec n\,, \qquad a=1,2\,,
	\end{equation}
	satisfying
	\begin{equation}
		\vec e_a\cdot \vec e_b = \delta_{ab}\,, \qquad \vec e_a\cdot \vec n = 0\,.
	\end{equation}
	Using the completeness relation
	\begin{equation}
		\label{completeness triad}
		\delta^{ij}=e^i_a\, \delta^{ab}\, e^j_b+n^i n^j\,,     
	\end{equation}
	we can decompose $O_i$ as
	\begin{equation}
		O^i=e^i_a\, O^a+n^i\, O\,, 
	\end{equation}
	written in terms of the $\operatorname{SO}(2)$ vector and scalar
	\begin{equation}
		O_a\equiv e^i_a\, O_i\,, \quad O\equiv n^iO_i\,.
	\end{equation}
	Projecting \eqref{carrollian action 1} onto these frame components yields, for the $\operatorname{SO}(2)$ vector $O_a$,
	\begin{equation}
		\label{carrollian action 2}
		\begin{split}
			[H, O_a]&=\partial_t O_a\,,\\
			[K,O_a]&=y^2\, \partial_t O_a\,,\\
			[B_a,O_b]&=y_a\, \partial_t O_b\,,\\
			[D,O_a]&=-i\left(t\partial_t + y^b\partial_b + \Delta\right)O_a\,,\\
			[P_a, O_b]&=- i\, \partial_a O_b\,,\\
			[K_a,O_b]&=\left(2 y^c[\Sigma_{ac}]+ 2iy_a\,(t\partial_t + y^c\partial_c + \Delta)- iy^2\partial_a\right)O_b\,,\\
			[L_{ab}\,,O_c] &= \left([\Sigma_{ab}]+ i\,(y_a\partial_b - y_b\partial_a)\right)O_c\,,
		\end{split}
	\end{equation}
	where $[\Sigma_{ab}]$ is the vector representation of the $\operatorname{SO}(2)$ generators $L_{ab}$, acting as
	\begin{equation}
		[\Sigma_{ab}]O_c = i\,(\delta_{ac}O_b - \delta_{bc}O_a)\,.
	\end{equation}
	For the scalar $O$ projection, the same $\mathfrak{iso}(1,3)$ action is found with $[\Sigma]=0$. As an illustration, let us compute $[D,O]=n^i[D,O_i]$ by projecting the corresponding equation in \eqref{carrollian action 1}. We explicitly have
	\begin{equation}
		\begin{split}
			n^i n^j [L_{3j}]^{(1)} O_i&=i\, n^i n^j(\delta_{i3} O_j-\delta_{ij} O_3)=i(n_3 O-O_3)=i\frac{2y^a}{1+y^2} O_a\,,\\
			n^i y^a \partial_a O_i&=y^a \partial_a(n^i O_i)-y^a \partial_a n^i\, O_i=y^a \partial_a O-\frac{2 y^a}{1+y^2} O_a\,,
		\end{split}
	\end{equation}
	such that the $\operatorname{SO}(3)$ spin term is canceled and we are left with
	\begin{equation}
		[D,O\,]=-i\left(t\partial_t + y^a\partial_a + \Delta\right)O\,.
	\end{equation}
	Hence the $\operatorname{SO}(3)$ vector field admits the simple decomposition
	\begin{equation}
		O_i \;\longrightarrow\; O_a \oplus O\,,
	\end{equation}
	i.e., the direct sum of two irreducible carrollian conformal fields of scaling dimension $\Delta$: one vector and one scalar. 
	
	\paragraph{General integer spin $s$.} The pattern seen above for the vector representation is just a special instance of the $\operatorname{SO}(3) \downarrow \operatorname{SO}(2)$ branching rule 
	\begin{equation}
		O_{i_1\dots\, i_s} \quad \longrightarrow\quad \bigoplus_{s'=0}^{s} O_{a_1\dots\, a_{s'}}\,,
	\end{equation}
	written in terms of totally symmetric and traceless tensor representations. In practice, each $\operatorname{SO}(2)$ irreducible component is obtained by projection onto
	the adapted frame $(\vec e_a,\vec n)$, followed by removal of the traces in the
	$\operatorname{SO}(2)$ indices, i.e.,
	\begin{equation}
		\label{irreducible components}
		O_{a_1\dots\, a_{s'}}=e^{i_1}_{\langle a_1}\dots e^{i_{s'}}_{a_{s'}\rangle}\,
		n^{i_{s'+1}}\dots n^{i_s}\, O_{i_1\dots \, i_s}\,, \qquad (s' \leq s)\,,
	\end{equation}
	where $\langle a_1\dots a_{s'}\rangle$ denotes the totally symmetric and traceless part with
	respect to $\delta_{ab}$. The traces that are removed carry no independent information: using
	the completeness relation \eqref{completeness triad} and the tracelessness of $O_{i_1\dots i_s}$,
	one has $\delta^{a_1a_2}e^{i_1}_{a_1}e^{i_2}_{a_2}=\delta^{i_1i_2}-n^{i_1}n^{i_2}$, such that each
	trace reduces to a frame component with two more indices along $\vec n$. 
	The proof that these projected components do indeed transform like irreducible carrollian conformal fields is given in Appendix~\ref{appendix general spin}. These irreducible carrollian conformal fields are in one-to-one correspondence with the massless particle states \cite{Nguyen:2023vfz}.
	
	Thus, the way we implement the flat limit contraction yields 
	\begin{equation}
		\mathcal{D}(\Delta,s) \qquad \stackrel{c \to 0}{\longrightarrow}\qquad \bigoplus_{s'=0}^s\, \mathcal U_{s'}\,,
	\end{equation}
	i.e., unitary conformal representations above the unitarity bound contract to a direct sum of massless Poincaré UIRs.  
	To our knowledge, this contraction had not been realized before directly at the level of the Hilbert space. The fate of the inner product will be discussed in Section~\ref{section inner product}, which will complement this statement. 
	
	\paragraph{Helicity basis.}  It is customary to label massless particles and the corresponding carrollian conformal fields with the Lorentz-invariant helicity $J$ rather than with $\operatorname{SO}(2)$ tensor indices, where the helicity basis $O_J$ satisfies 
	\begin{equation}
		[\Sigma_{ab}] O_J=J \epsilon_{ab}\, O_J\,.
	\end{equation}
	Obviously, the scalar $O$ has helicity $J=0$. The $J=\pm 1$ helicity modes $O_\pm$ of the vector representation $O_a$ are simply given by
	\begin{equation}
		O_\pm = O_1\pm i O_2\,,
	\end{equation}
	as may be explicitly checked upon acting with $[\Sigma_{ab}]$. These may be constructed directly from the $\operatorname{SO}(3)$ vector $O_i$ by projecting onto the complex triad,
	\begin{equation}
		m^i=e^i_1+i e^i_2\,, \qquad \bar m^i=e^i_1-i e^i_2\,, \qquad n^i\,,
	\end{equation}
	satisfying
	\begin{equation} 
		m\cdot m = 0\,, \qquad m\cdot\bar m = 2\,, \qquad m\cdot n = 0\,.
	\end{equation}
	For any totally symmetric traceless tensor field $O_I=O_{i_1...\,i_s}$, the $(2s+1)$ helicity components are explicitly given by
	\begin{equation}
		O_J \equiv
		\begin{cases}
			m^{i_1}\dots m^{i_J}\, n^{i_{J+1}} \dots n^{i_s}\, O_{i_1\dots\, i_s}\,, \quad & J = 0, 1, \dots, s\,,\\
			\bar m^{i_1}\dots \bar m^{i_{|J|}}\, n^{i_{|J|+1}}\dots n^{i_s}\, O_{i_1\dots\, i_s}\,, \quad & J = -1, \dots, -s\,.
		\end{cases}
	\end{equation}

	\subsection{Shortening condition}
	\label{section 3.4}
	The discussion of the flat limit contraction to massless particle representations was so far implemented for generic scaling dimension above the unitarity bound. Precisely at the bound $\Delta=s+1$ for $s \geq 1$, unitary conformal representations are short and are described in terms of conserved currents. Let us see how this affects the flat limit contraction, by following the shortening condition \eqref{conservation equation} through the various steps described above.
	
	We first express the shortening condition \eqref{conservation equation} in radial coordinates using \eqref{Jacobian derivatives}, which yields
	\begin{equation}
		\label{SC-radial}
		\partial_r\left(n^i  O_{i i_2\dots i_s}\right)
		+ \frac{1}{r} D^i O_{i i_2\dots i_s} = 0\,.
	\end{equation}
	Using \eqref{r tau relation}, this is equivalent to
	\begin{equation}
		\label{SC-tau}
		\partial_\tau \left(n^i  O_{i i_2\dots i_s}\right)
		= \frac{4c}{4 - c^2\tau^2}\; D^i O_{i i_2\dots i_s}\,.
	\end{equation}
	Therefore, in the flat limit contraction $c \to 0$ at fixed $\tau$, this reduces to the simple constraint
	\begin{equation}
		\label{SC-limit}
		\partial_\tau\left(n^i O_{i i_2\dots i_s}(\tau,\vec n)\right) = 0\,.
	\end{equation}
	In the flat conformal frame defined by \eqref{flat conformal frame}, this also reads
	\begin{equation}
		\label{SC-limit-2}
		\partial_t \big(n^i \hat O_{i i_2\dots i_s}(t,y)\big) = 0\,.
	\end{equation}
	In terms of the irreducible components \eqref{irreducible components}, only the spin-$s$ field $\hat O_{a_1\dots\, a_s}$ remains unconstrained, while all other components are required to be time-independent. We will argue in Section~\ref{section inner product} that this actually completely removes all these lower spin components, such that the flat limit contraction yields a single massless particle of spin $s$,
	\begin{equation}
		\label{contraction short multiplet}
		\mathcal{D}(s+1,s) \qquad \stackrel{c \to 0}{\longrightarrow}\qquad \mathcal U_s= \mathcal U^{(-s)} \oplus \mathcal U^{(s)}\,,
	\end{equation}
	where the last expression displays the split of the spin-$s$ tensor representation $\mathcal{U}_s$ into two independent helicity components $\mathcal U^{(J)}$ with $J=\pm s$, paired under parity.
	
	\section{Inner product in the flat limit}
	\label{section inner product}
	
	We have demonstrated how massless particle representations are obtained from conformal irreps in the flat limit contraction $\mathfrak{so}(2,3) \to \mathfrak{iso}(1,3)$. To complete the discussion, we now show how the inner product \eqref{state overlap} reduces to the Lorentz-invariant inner
	product of massless particles.
	
	\subsection{Zero spin}
	\label{section 4.1}
	We start with the scalar case $s=0$ for simplicity.
	In the coordinates \eqref{momentum rapidity param}, the overlap of pseudo-momentum states
	\eqref{pseudo-momentum kernel} reads
	\begin{equation}
		\label{scalar overlap rapidity}
		\langle p_1|p_2\rangle
		= \left(\cosh^2 \left(\frac{\eta_1-\eta_2}{2}\right) + \frac14\sinh\eta_1\sinh\eta_2\,|\vec n_{12}|^2\right)^{-\Delta}\,.
	\end{equation}
	Changing to the variable $\tau$ defined in \eqref{tau definition}, this becomes
	\begin{equation}
		\label{scalar overlap tau exact}
		\langle p_1|p_2\rangle
		=\left( \frac{4c^2\,\bar \tau_1\tau_2}
		{c^2(\bar \tau_1+\tau_2)^2 + \big(1-\frac14c^2\bar \tau_1^2\big)\big(1-\frac14c^2\tau_2^2\big)\,|\vec n_{12}|^2}\right)^\Delta\,,
	\end{equation}
	where the complex conjugate $\bar \tau_1$ appears so as to accommodate the Wick rotation $\tau=-iu+0^+$ encountered in Section~\ref{section 3.1}.
	In the limit $c\to0$ at fixed $\tau$, this overlap vanishes as $c^{2\Delta}$ at any fixed separation
	$\vec n_{12}\neq0$, while it remains of order one for $\vec n_{12}\sim c$. Thus, in the limit $c \to 0$, the distribution \eqref{scalar overlap tau exact} is fully supported at $\vec n_1=\vec n_2$. More precisely, we obtain
	\begin{equation}
		\label{flat limit inner product approx}
		\langle p_1|p_2\rangle
		\approx c^2\, \frac{\pi}{(\Delta-1)}\frac{(4\bar \tau_1\tau_2)^\Delta}{(\bar \tau_1+\tau_2)^{2(\Delta-1)}}\, \delta^{(2)}(\vec n_1,\vec n_2)\,, \qquad (c \to 0)\,,
	\end{equation}
	where $\delta^{(2)}(\vec n_1,\vec n_2)$ is the invariant delta function on $S^2$ satisfying
	\begin{equation}
		\int d\Omega_1\, f(\vec n_1)\, \delta^{(2)}(\vec n_1,\vec n_2)=f(\vec n_2)\,, \qquad \forall \, f(\vec n)\,.
	\end{equation}
	To demonstrate this, we first integrate \eqref{scalar overlap tau exact} over the sphere coordinate $\vec n_1 \in S^2$,
	keeping $\vec n_2$ fixed.
	To this end, we adopt a system of polar angles centered around $\vec n_2$,
	\begin{equation}
		\label{polar coordinates}
		d\Omega_1=\sin \theta\, d\theta\, d\varphi = |\vec n_{12}|\, d|\vec n_{12}|\, d\varphi \,,
	\end{equation}
	where $|\vec n_{12}|=2\sin \tfrac{\theta}{2} \in [0,2]$ is the chordal distance between $\vec n_1$ and $\vec n_2$. Introducing the rescaled variable $\rho=|\vec n_{12}|/c$, we therefore have
	\begin{equation}
		\begin{split}
			\int d\Omega_1\, \langle p_1|p_2\rangle&=2\pi c^2 \int_0^{2/c} d\rho\, \rho\,  \langle p_1|p_2\rangle\,.
		\end{split}
	\end{equation}
	In the limit $c \to 0$ and provided $\Delta>1$, this integral converges to
	\begin{equation}
		\label{angular integral scalar}
		\begin{split}
			\lim_{c \to 0} c^{-2} \int d \Omega_1\, \langle p_1|p_2\rangle&=2\pi \int_0^\infty d\rho\, \rho\left( \frac{4\bar \tau_1\tau_2}
			{(\bar \tau_1+\tau_2)^2 + \rho^2}\right)^\Delta=\frac{\pi}{(\Delta-1)} \frac{(4\bar \tau_1\tau_2)^\Delta}{(\bar \tau_1+\tau_2)^{2(\Delta-1)}}\,,
		\end{split}
	\end{equation}
	in agreement with \eqref{flat limit inner product approx}.
	
	It is therefore natural to equip the states $|\tau,\vec n\rangle$ defined in \eqref{states tau definition} with a \textit{renormalized} inner product that stays well-defined in this limit, namely
	\begin{equation}
		\label{tau overlap}
		\langle \tau_1,\vec n_1|\tau_2,\vec n_2\rangle\equiv \lim_{c \to 0} c^{-2}\, (4\bar \tau_1 \tau_2)^{-\Delta} \langle p_1|p_2\rangle=\frac{\pi}{(\Delta-1)}\frac{\delta^{(2)}(\vec n_1,\vec n_2)}{(\bar \tau_1+\tau_2)^{2(\Delta-1)}} \,,
	\end{equation}
	or equivalently
	\begin{equation}
		\label{carrollian inner product}
		\langle u_1,\vec n_1|u_2,\vec n_2\rangle=\frac{\pi}{(\Delta-1)}\frac{\delta^{(2)}(\vec n_1,\vec n_2)}{(i u_{12}+0^+)^{2(\Delta-1)}}\,, \qquad u_{12}\equiv u_1-u_2\,.
	\end{equation}
	It may be checked that \eqref{carrollian inner product} is the carrollian version of the Lorentz-invariant inner product of massless momentum states, usually given by the standard formula
	\begin{equation}
		\label{momentum inner product}
		\langle q_1|q_2\rangle = (2\pi)^3 \,2\omega_1\,\delta^{(3)}(\vec q_1-\vec q_2)
		= 2(2\pi)^3\,(\omega_1)^{-1}\, \delta(\omega_1-\omega_2)\,\delta^{(2)}(\vec n_1,\vec n_2)\,.
	\end{equation}
	Indeed, we can also recover \eqref{tau overlap} directly from \eqref{carrollian states} and \eqref{momentum inner product},
	\begin{equation}
		\begin{aligned}
			\langle \tau_1,\vec n_1|\tau_2,\vec n_2\rangle
			&= |\mathcal N|^2\int_0^\infty d\omega_1\,d\omega_2\,(\omega_1\omega_2)^{\Delta-1}\,
			e^{-\omega_1\bar\tau_1-\omega_2\tau_2}\,\langle q_1|q_2\rangle \\
			&= 2(2\pi)^3\,|\mathcal N|^2\,\delta^{(2)}(\vec n_1,\vec n_2)
			\int_0^\infty d\omega\;\omega^{2\Delta-3}\,e^{-\omega(\bar\tau_1+\tau_2)} \\
			&= 2(2\pi)^3\,|\mathcal N|^2\,\Gamma(2\Delta-2)\,
			\frac{\delta^{(2)}(\vec n_1,\vec n_2)}{(\bar\tau_1+\tau_2)^{2(\Delta-1)}}\,,
		\end{aligned}
	\end{equation}
	which agrees with \eqref{tau overlap} for the normalization
	\begin{equation}
		\label{scalar normalization}
		|\mathcal{N}|^2=\frac{\pi}{2(2\pi)^3 (\Delta-1)\Gamma(2\Delta-2)}=\frac{1}{8\pi^2\,\Gamma(2\Delta-1)}\,.
	\end{equation}
	We note that the carrollian inner product \eqref{carrollian inner product} is singular at $\Delta=1$, which is a well-known feature in the Carrollian Holography literature \cite{Donnay:2022wvx,Nguyen:2023miw,Nguyen:2025zhg,Nguyen:2025sqk}.
	
	\subsection{Integer spin}
	\label{section 4.2}
	We now extend the analysis of the flat limit of the inner product carried out in Section~\ref{section 4.1} to arbitrary integer spin. In doing so, we will also establish that the components $O_J$ with $|J|<s$
	become null at the unitarity bound $\Delta=s+1$ and are therefore discarded, as anticipated in Section~\ref{section 3.4}.
	
	In the radial foliation \eqref{radial foliation}, the inner product \eqref{state overlap} becomes
	\begin{equation}
		\label{IP radial}
		\langle O_I(r_1,\vec n_1)| O_J(r_2,\vec n_2)\rangle
		= \frac{\mathcal I\indices{_I^K}(r_1 \vec n_1)\,\mathcal I_{KJ}(\vec n_1/r_1 -r_2\,\vec n_2)}{\left[(1-r_1r_2)^2 + r_1r_2\,|\vec n_{12}|^2\right]^\Delta}\,,
		\qquad \vec n_{12}\equiv\vec n_1-\vec n_2\,.
	\end{equation} 
	Applying the change of coordinate \eqref{r tau relation}, the scalar factor in \eqref{IP radial}
	multiplied by the conformal factors arising from \eqref{Psi(p) from O(x)} is precisely the scalar
	overlap \eqref{scalar overlap tau exact}, whose flat limit has already been worked out in
	Section~\ref{section 4.1}.
	As demonstrated in Appendix~B of \cite{Nguyen:2026rwt}, the product of inversion tensors is a spatial rotation $\mathcal R_\theta$ around $\vec n_1 \times \vec n_2$ by an angle $\theta$ determined by
	\begin{equation}
		\tan \frac{\theta}{2}=\frac{r_1 r_2 |\vec n_1 \times \vec n_2|}{1-r_1 r_2\, (\vec n_1 \cdot \vec n_2)}\,.
	\end{equation}
	As we have seen in Section~\ref{section 4.1}, the scalar factor localizes the overlap on a vanishingly small region $|\vec n_{12}|\sim c$, which we resolved with the polar coordinates
	\eqref{polar coordinates} centered at $\vec n_2$ and the rescaled chordal distance
	$\rho=|\vec n_{12}|/c$.
	Using
	\begin{equation}
		\begin{split}
			1-r_1r_2\,(\vec n_1\cdot\vec n_2) &= (1-r_1r_2) + \tfrac12\, r_1r_2\,|\vec n_{12}|^2\,,\\
			|\vec n_1\times\vec n_2| &= |\vec n_{12}|\,\sqrt{1-\tfrac14|\vec n_{12}|^2}\,,
		\end{split}
	\end{equation}
	together with $1-r_1r_2 = c\,(\bar \tau_1+\tau_2)+O(c^2)$ which follows from \eqref{r tau relation}, the rotation angle simplifies to
	\begin{equation}
		\tan \frac{\theta}{2}=\frac{\rho}{( \bar \tau_1+\tau_2)}+O(c)\,.
	\end{equation}
	Since rotation matrices are bounded, the argument leading to \eqref{tau overlap}
	goes through unchanged, the angular integral now being weighted by the rotation $[\mathcal R_\theta]_{IJ}$.
	Thus, the renormalized
	inner product of the states \eqref{states tau definition} becomes
	\begin{equation}\label{IP-tau}
		\begin{split}
			{}_I\langle \tau_1,\vec n_1|\tau_2,\vec n_2\rangle_J
			&\equiv \lim_{c \to 0} c^{-2}\, (4 \bar \tau_1 \tau_2)^{-\Delta}\; {}_I\langle p_1|p_2\rangle_J\\
			&=\delta^{(2)}(\vec n_1,\vec n_2)
			\int_0^{\infty} d\rho\, \rho \int_0^{2\pi} d\varphi\, \frac{[\mathcal R_{\theta}]_{IJ}}{\left((\bar  \tau_1+\tau_2)^2+\rho^2\right)^\Delta}\\
			&=\frac{\delta^{(2)}(\vec n_1,\vec n_2)}{( \bar \tau_1+\tau_2)^{2(\Delta-1)}}
			\int_0^{\infty} dw\, w \int_0^{2\pi} d\varphi\, \frac{[\mathcal R_{\theta}]_{IJ}}{\left(1+w^2\right)^\Delta}\,,
		\end{split}
	\end{equation}
	where in the last equality we changed variable $w=\rho/(\bar \tau_1+\tau_2)$, such that
	$\tan \tfrac{\theta}{2}=w$.
	The angular integral 
	\begin{equation}
		K_{IJ}\equiv \int_0^{\infty} dw\, w \int_0^{2\pi} d\varphi\, \frac{[\mathcal R_{\theta}]_{IJ}}{\left(1+w^2\right)^\Delta}\,,
	\end{equation}
	converges for $\Delta>1$, which is automatically satisfied for $s\geq1$ by the
	unitarity bounds \eqref{unitarity bounds}. 
	
\begin{figure}[t]
		\centering
		\tdplotsetmaincoords{60}{160}
		\begin{tikzpicture}[tdplot_main_coords,>=Stealth,scale=1.6,
			vec/.style={->,thick},
			lab/.style={font=\small}]
			\shade[tdplot_screen_coords,ball color=gray!10,opacity=0.6] (0,0,0) circle (2.0);
			\draw[tdplot_screen_coords,gray] (0,0,0) circle (2.0);
			\tdplotdrawarc[gray!60,dashed]{(0,0,0)}{2.0}{0}{360}{}{}
			\fill (0,0,0) circle (0.6pt);
			\draw[gray,thin,dashed] (0,0,0) -- (1.6383,0.0000,1.1472);
			\draw[gray,thin,dashed] (0,0,0) -- (1.8023,0.4064,0.7659);
			\filldraw[fill=blue!8,draw=blue!40,opacity=0.85] (0.9213,-1.2500,2.1711) -- (2.3553,-1.2500,0.1232) -- (2.3553,1.2500,0.1232) -- (0.9213,1.2500,2.1711) -- cycle;
			\draw[vec,blue!70!black] (1.6383,0.0000,1.1472) -- (2.3553,0.0000,0.1232) node[lab,anchor=north] {$\vec e_1$};
			\draw[vec,blue!70!black] (1.6383,0.0000,1.1472) -- (1.6383,1.2500,1.1472) node[lab,anchor=north] {$\vec e_2$};
			\draw[vec,blue!70!black] (1.6383,0.0000,1.1472) -- (2.6622,0.0000,1.8641) node[lab,anchor=south] {$\vec n_2$};
			\draw[vec,red!75!black] (1.6383,0.0000,1.1472) -- (1.8023,0.4064,0.7659);
			\fill[red!75!black] (1.8023,0.4064,0.7659) circle (1.2pt) node[lab,anchor=north] {$\vec n_{12}$};
			\fill[blue!70!black] (1.6383,0.0000,1.1472) circle (1.2pt);
		\end{tikzpicture}
		\caption{Geometric setup used for the flat limit of the inner product. In the polar
			coordinates centered at $\vec n_2$, the chord $\vec n_{12}$ makes an angle $\varphi$ with
			$\vec e_1$ in the plane tangent to the sphere, spanned by the adapted triad vectors
			$\vec e_1,\vec e_2$.}
		\label{fig:triad}
	\end{figure}
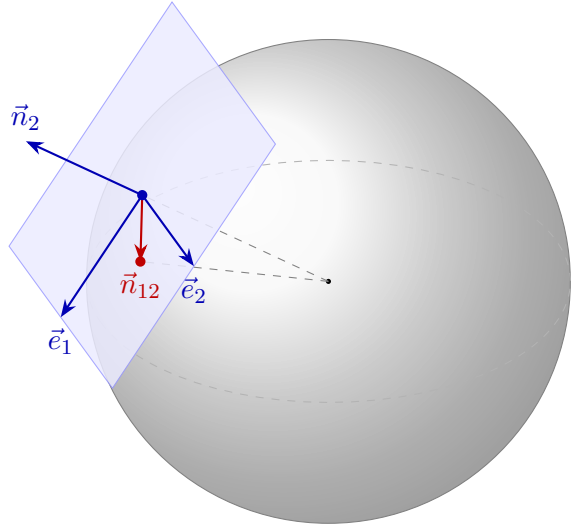

	\paragraph{Adapted triad.}
	We wish to show that $K_{IJ}$ does not mix massless states of different helicity. This becomes transparent in the adapted frame components
	$O_{A_1\dots\, A_s}$ introduced in Section~\ref{section 3.3}, obtained by contracting each index
	with the orthogonal matrix $R\indices{_A^i}$ with components
	\begin{equation}
		R\indices{_a^i}=e^i_a\,, \qquad R\indices{_3^i}=n^i\,.
	\end{equation}
	We contract the bra with
	the triad at $\vec n_1$ and the ket with the triad at $\vec n_2$. Since the overlap is supported
	on $|\vec n_{12}|\sim c$, the two triads coincide in the limit, and both can be evaluated at
	$\vec n_2$. We also adopt again the polar coordinates \eqref{polar coordinates} with $\varphi$ chosen such that the direction of $\vec n_{12}$ is $\cos\varphi\,\vec e_1+\sin\varphi\,\vec e_2$ in the limit $c \to 0$ ; see Figure~\ref{fig:triad}. The rotation axis then becomes
	\begin{equation}
		\vec n_1\times\vec n_2
		\;\stackrel{c \to 0}{\longrightarrow}\; \left(\cos\varphi\,\vec e_1+\sin\varphi\,\vec e_2\right) \times \vec n_2
		= \sin\varphi\, \vec e_1-\cos\varphi\, \vec e_2\,,
	\end{equation}
	where we used $\vec n_2 \times\vec e_1=\vec e_2$ and $\vec n_2 \times\vec e_2=-\vec e_1$. This is the second axis of the triad rotated by $\varphi$ about the third axis, with reversed orientation,
	\begin{equation}
		\sin\varphi\,\vec e_1-\cos\varphi\,\vec e_2=-\mathcal R_3(\varphi)\,\vec e_2\,.
	\end{equation}
	In adapted frame components, the limiting rotation $\mathcal R_\theta$ is therefore a rotation by $-\theta$ about the second axis,
	conjugated by a rotation by $\varphi$ about the third axis,
	\begin{equation}
		\label{rotation in triad}
		R\,\mathcal R_\theta\, R^T = \mathcal R_3(\varphi)\, \mathcal R_2(-\theta)\, \mathcal R_3(-\varphi)\,,
	\end{equation}
	where conjugation by $R$ has removed the explicit dependence on $\vec n_2$.
	
	The inner product in adapted frame components therefore reads
	\begin{equation}
		\label{IP-triad}
		{}_A\langle \tau_1,\vec n_1|\tau_2,\vec n_2\rangle_B
		= K_{AB}\;\frac{\delta^{(2)}(\vec n_1,\vec n_2)}{(\bar \tau_1+\tau_2)^{2(\Delta-1)}}\,,
	\end{equation}
	with
	\begin{equation}
		\label{K triad}
		K_{AB}=\int_0^\infty  \frac{dw\, w}{(1+w^2)^\Delta} \int_0^{2\pi} d\varphi\;
		\big[\mathcal R_3(\varphi)\,\mathcal R_2(-\theta)\,\mathcal R_3(-\varphi)\big]_{AB}\,,
	\end{equation}
	where $A=A_1\dots A_s$ and $B=B_1\dots B_s$ are multi-indices of the adapted frame, and rotations act in the rank-$s$ tensor
	representation. The integral over $\varphi$ averages over rotations about the third axis, such that
	$K_{AB}$ commutes with them. In other words, $K_{AB}$ is an $\operatorname{SO}(2)$-invariant
	bilinear form on totally symmetric traceless tensors, so that it does not mix the irreducible components
	\eqref{irreducible components}. 
	
	\paragraph{Helicity components.}
	Starting from the adapted frame components, the helicity components defined in  Section~\ref{section 3.3} are obtained by contracting with the complex triad $m^A=(1,i,0)$ and $n^A=(0,0,1)$, namely
	\begin{equation}
		|\tau,\vec n\rangle_J \equiv
		\begin{cases}
			m^{A_1}\dots m^{A_J}\, n^{A_{J+1}} \dots  n^{A_s}\, |\tau,\vec n\rangle_{A_1\dots\, A_s}\,, \quad & J = 0, 1, \dots, s\,,\\
			\bar m^{A_1}\dots \bar m^{A_{|J|}}\, n^{A_{|J|+1}}\dots  n^{A_s}\, |\tau,\vec n\rangle_{A_1\dots\, A_s}\,, \quad & J = -1, \dots, -s\,,
		\end{cases}
	\end{equation}
	with the dual bras contracted with the complex conjugate vectors. Since the frame
	components are totally symmetric and traceless, only the traceless projection $v_J$ of
	$m^{\otimes|J|}\otimes n^{\otimes(s-|J|)}$, respectively of
	$\bar m^{\otimes|J|}\otimes n^{\otimes(s-|J|)}$, contributes. It is an eigenvector of
	rotations about the third axis,
	\begin{equation}
		\mathcal R_3(\varphi)\, v_J = e^{-iJ\varphi}\, v_J\,,
	\end{equation}
	with norm
	\begin{equation}
		\label{NsJ}
		N_{s,J}^2 \equiv \bar v_J\cdot v_J = \frac{2^s(s+|J|)!\,(s-|J|)!}{(2s)!}\,.
	\end{equation}
	Evaluating \eqref{IP-triad}-\eqref{K triad} in this helicity basis, the dependence on $\varphi$ reduces to the phase
	$e^{i(J_1-J_2)\varphi}$, such that the average over $\varphi$ enforces $J_1=J_2$. The
	remaining matrix element of $\mathcal R_2(-\theta)$ between normalized helicity vectors is the
	diagonal Wigner function $d_{JJ}^s(\theta)$ \cite{Edmonds:1957,Varshalovich:1988ifq},
	\begin{equation}
		\bar v_J \cdot \mathcal R_2(-\theta)\, v_J=N_{s,J}^2\,  d^s_{JJ}(\theta)=N_{s,J}^2 \left(\cos \tfrac{\theta}{2}\right)^{2|J|} P^{(0,\,2|J|)}_{s-|J|}(\cos \theta)\,,
	\end{equation}
	expressed in terms of the Jacobi polynomial $P^{(\alpha,\beta)}_\ell(x)$. Changing variable $z=\cos \tfrac{\theta}{2}$, the relevant helicity component of \eqref{K triad} becomes
	\begin{equation}
		\label{KJJ}
		\begin{split}
			K_{JJ}\equiv \bar v_J \cdot K\, v_J&=2\pi N_{s,J}^2 \int_0^1 dz\, z^{2\Delta-3} d^s_{JJ}(2\arccos z)\\
			&=2\pi N_{s,J}^2 \int_0^1 dz\, z^{2(\Delta+|J|)-3}\, P^{(0,\,2|J|)}_{s-|J|}(2z^2-1)\\
			&=\pi N_{s,J}^2\, \frac{\Gamma(\Delta-|J|-1)\,\Gamma(\Delta+|J|-1)}{\Gamma(\Delta-s-1)\,\Gamma(\Delta+s)}\,,
		\end{split}
	\end{equation}
	such that
	\begin{equation}
		\label{IP-helicity}
		{}_{J_1}\langle \tau_1,\vec n_1|\tau_2,\vec n_2\rangle_{J_2}
		= \delta_{J_1J_2}\, K_{J_1J_1}\,
		\frac{\delta^{(2)}(\vec n_1,\vec n_2)}{(\bar \tau_1+\tau_2)^{2(\Delta-1)}}\,.
	\end{equation}
	Different helicities are therefore orthogonal, with the same carrollian
	two-point structure found for $s=0$.
	Note that one has $N_{0,0}=1$ and $K_{00}=\pi/(\Delta-1)$, in agreement with
	\eqref{tau overlap}.
	
	\paragraph{Flat conformal frame.}
	Since $K_{AB}$ is a constant matrix, the conformal transformation \eqref{flat conformal frame} only
	affects the scalar prefactor. All powers of the conformal factor $\Omega$ cancel, such that we simply have
	\begin{equation}\label{IP-flat}
		\langle \hat O_{A}(t_1,y_1)|\hat O_{B}(t_2,y_2)\rangle
		= \frac{K_{AB}}{2}\;\frac{\delta^{(2)}(y_{12})}{\left(\bar t_1+t_2\right)^{2(\Delta-1)}}=\frac{K_{AB}}{2}\;\frac{\delta^{(2)}(y_{12})}{(i u_{12}+0^+)^{2(\Delta-1)}}\,,
	\end{equation}
	where in the first equality we allowed for complex $t_1, t_2$, and where in the second equality we implemented the Wick rotation $t=-iu+0^+$. The inner product \eqref{IP-flat}
	reproduces the standard carrollian two-point function, modulo a helicity-dependent normalization factor.
	
	\paragraph{Momentum basis.}
	Each helicity component is related to massless momentum eigenstates $|q\rangle_J$ of helicity
	$J$ exactly as in the scalar case,
	\begin{equation}
		|\tau,\vec n\rangle_J =\mathcal{N}_J\, \int_0^\infty d\omega\, \omega^{\Delta-1} e^{-\omega \tau}\, |q(\omega,\vec n)\rangle _J\,,
	\end{equation}
	since the derivation leading to \eqref{carrollian states} only involves the action of
	$\mathcal P_\mu$ and of the boost along $\vec n$, both of which act diagonally on the helicity
	components. Repeating the computation leading to \eqref{tau overlap}, the Lorentz-invariant
	inner product \eqref{momentum inner product}, now supplemented with $\delta_{J_1J_2}$, is
	recovered for
	\begin{equation}
		\label{NJ}
		|\mathcal N_J|^2=\frac{K_{JJ}}{2(2\pi)^3\,\Gamma(2\Delta-2)}\,,
	\end{equation}
	which reduces to \eqref{scalar normalization} for $s=0$.
	
	\subsection{Short multiplets and null states}
	\label{section 4.3}
	At the unitarity bound $\Delta=s+1$, the gamma function $\Gamma(\Delta-s-1)$ in the denominator of \eqref{KJJ} makes the
	coefficients $K_{JJ}$ vanish for $|J|<s$. For $J=\pm s$, it is canceled by the gamma function $\Gamma(\Delta-|J|-1)$ in the numerator, such that
	\begin{equation}
		K_{JJ}=\frac{\pi N^2_{s,s}}{2s}=\frac{2^{s-1}\pi}{s}\,, \qquad J=\pm s\,.
	\end{equation}
	Hence all
	components $O_J$ with $|J|<s$, or equivalently all frame components with at least one index along
	$\vec n$, have vanishing norm,
	\begin{equation}
		n^{A_1} K_{A_1\dots } = 0\,,
	\end{equation}
	and must be removed from the physical spectrum.
	This is the inner-product counterpart of the shortening condition \eqref{SC-limit-2}. These
	null components decouple from the physical Hilbert space, and only the massless particles of maximal helicity $J=\pm s$ survive,
	which establishes \eqref{contraction short multiplet} as anticipated in Section~\ref{section 3.4}. Equivalently,
	$\mathcal N_J=0$ for $|J|<s$ in \eqref{NJ}. 
	
	\section{Discussion}
	\label{section discussion}
	
	In this paper, we presented a mechanism by which massless scattering states emerge from unitary lowest-weight representations $\mathcal D(\Delta,s)$ of the conformal group, in the flat limit of AdS$_4$ described by the Inönü--Wigner contraction $\mathfrak{so}(2,3) \to \mathfrak{iso}(1,3)$. We showed that this contraction can be implemented such that unitary representations $\mathcal D(\Delta,s)$, with fixed $\Delta$ and arbitrary integer spin $s$, contract to unitary Poincaré representations describing massless particles. For $\Delta>s+1$, the result is the reducible direct sum \eqref{result generic} of massless particles of all spins $s'\leq s$. At the unitarity bound, the result is a single massless particle \eqref{result short} of spin $s$, including both helicities $J=\pm s$. The short multiplets $\mathcal D(s+1,s)$, singled out historically by gauge invariance \cite{Fronsdal:1975eq} and by their conformal extension \cite{Angelopoulos:1980wg}, are precisely those that contract to a single massless particle. In contrast with the contraction studied in \cite{Angelopoulos:1980wg}, that yields a single helicity, ours produces both helicities $J=\pm s$. It thereby accounts for the reduction of spin-$s$ gauge fields in AdS$_4$ to spin-$s$ gauge fields in Minkowski spacetime. Both results \eqref{result generic}--\eqref{result short} realize the branching rule $\operatorname{SO}(3)\downarrow\operatorname{SO}(2)$ identified in \cite{Brink:2000ag,Basile:2016aen}, with the lower spins projected out at the unitarity bound. 
	
	These results rest on implementing the Inönü--Wigner contraction \textit{simultaneously} with an infinite boost on the pseudo-momentum states first introduced by Fronsdal \cite{Fronsdal:1965zzb,Fronsdal:1974ew} and further developed in \cite{Nguyen:2026rwt}, in line with the suggestion of Berenstein and Simon \cite{Berenstein:2025tts}. We provided an equivalent description in terms of local conformal fields in $\mathbb{R}^3$ familiar from radial quantization, wherein massless scattering states arise from operator insertions in a vanishingly small neighborhood of the unit sphere; see Figure~\ref{fig:ball to hyperboloid}. Unlike massive particle states that emerge directly in the momentum basis \cite{Nguyen:2026rwt}, massless scattering states naturally emerge in a \textit{carrollian} basis of scaling dimension $\Delta$, where energy $\omega$ is traded for retarded time $u$ at null infinity after the Wick rotation $\tau=-iu+0^+$. This carrollian basis is the preferred one from the viewpoint of Carrollian Holography \cite{Banerjee:2018gce,Banerjee:2019prz,Bagchi:2022emh,Donnay:2022wvx,Nguyen:2023vfz,Nguyen:2025zhg}. This also explains why the massless wavefunctions found in \cite{Berenstein:2025tts} are not momentum eigenstates, and why they retain the memory of $\Delta$: in the flat limit, $\Delta$ becomes the scaling dimension labeling the carrollian basis, rather than a label of the Poincaré representation itself.\footnote{The Lorentz content of a massless particle UIR is uniquely fixed and only involves
		representations of the principal continuous series \cite{Iacobacci:2024laa}, so that the real
		dimension $\Delta$ inherited from $\mathcal D(\Delta,s)$ cannot label a Lorentz representation
		carried by $\mathcal U_s$; it labels instead the weight $\omega^{\Delta-1}$ of the transform
		\eqref{carrollian states}.} After a single, helicity-independent rescaling, the inner product of $\mathcal D(\Delta,s)$ converges to the Poincaré-invariant inner product in its carrollian form \eqref{IP-helicity}. Each helicity $|J|\leq s$ has a finite and positive norm \eqref{KJJ} for $\Delta>s+1$. At the unitarity bound, the norms of the lower helicities $|J|<s$ vanish, consistently with the shortening condition \eqref{SC-limit}. 
	
	We now discuss these results in relation with various other approaches to the flat limit of AdS/CFT, and comment on some open questions.
	
	\paragraph{Longitudinal polarizations.}
	In extending the bulk wavefunction analysis of \cite{Berenstein:2025tts} to spinning fields, it was found that longitudinal polarizations acquire an additional diverging factor $e^\eta \to \infty$ in the flat limit \cite{Berenstein:2025qhb}. The authors suggested that this divergence should be cured by treating the longitudinal modes as Goldstone fields. 
	
	The analysis presented here confirms this expectation and clarifies its origin. This longitudinal enhancement is the AdS counterpart of the familiar growth $\epsilon_L^\mu\simeq p^\mu/m$ of longitudinal polarization vectors for massive fields in the high-energy limit \cite{Chanowitz:1985hj,Veltman:1989ud}, already discussed in the context of the infinite boost limit of massive particles in \cite{Nguyen:2026yho}. 
	The divergence therefore resides in the polarization tensors intertwining covariant fields with particle states, while the states themselves have finite norms.
	The lower-helicity particles in \eqref{result generic} are those that a covariant description assigns to the Stückelberg fields, and the equivalence theorem \cite{Chanowitz:1985hj,Veltman:1989ud} suggests that amplitudes of longitudinal modes reduce to those of the corresponding Goldstone particles. For $\Delta=s+1$, the lower-helicity components have zero norm, consistently with the observation in \cite{Berenstein:2025qhb} that longitudinal modes are pure gauge for massless AdS fields. 
	
	\paragraph{Boundary operators and mode functions.}
	Most flat limit prescriptions construct flat space creation and annihilation operators out of Lorentzian boundary operators. In a global mode expansion, massless waves of energy $\omega$ arise from AdS mode solutions whose radial quantum number scales as $n\sim\ell\omega/2$ \cite{Fitzpatrick:2011jn,Li:2021snj}. Inverting this expansion expresses momentum space operators as CFT primaries smeared over a global time interval of vanishing width $\ell^{-1}$ around $\tau_{\text{AdS}}=\pm\tfrac{\pi}{2}$ \cite{Hijano:2019qmi,Li:2021snj}. For massive particles, the interval is shifted to complex times, in qualitative agreement with the insertion points of the pseudo-momentum states \eqref{Psi(p) from O(x)} being placed in $\mathbb{R}^3$. For massless particles at fixed $\Delta$, the smearing kernel reduces to a Fourier transform in retarded time, with normalization proportional to $\ell^{1-\Delta}\, \omega^{1-\Delta}$ \cite{Fitzpatrick:2011hu,Hijano:2019qmi,Li:2021snj}. This is precisely the inverse transform \eqref{inverse transform bis}, and we can identify these massless smearing prescriptions with Fourier transforms of the carrollian states $|u,\vec n\rangle$. For spinning fields, these constructions have mostly been restricted to spin-1 conserved currents \cite{Hijano:2020szl}, while the flat limit at fixed $\Delta$ of non-conserved spinning operators and the resulting lower-helicity massless particles had not been considered.
	
	\paragraph{Carrollian bases from bulk-to-boundary propagators.}
	For scalar fields, the emergence of a carrollian basis at fixed $\Delta$ has also been established at the level of bulk wavefunctions. De Gioia and Raclariu showed that AdS$_4$ bulk-to-boundary propagators with boundary insertions at global Lorentzian times $\tau_{\text{AdS}}=\pm\tfrac{\pi}{2}+\tfrac{u}{\ell}$ reduce in the flat limit to outgoing or incoming carrollian wavefunctions of scaling dimension $\Delta$ \cite{deGioia:2022fcn,deGioia:2024yne}. This agrees with the picture provided here, the only difference being that our insertions are initially Euclidean. They lie at a radial distance $1-r\simeq c\,\tau$ from the unit sphere, which plays the role of the time slices $\tau_{\text{AdS}}=\pm\tfrac{\pi}{2}$, and the Euclidean offset becomes the Lorentzian shift $u/\ell$ after the Wick rotation $\tau=-iu+0^+$. More recently, Navarro and Raclariu showed that time-ordered bulk-to-boundary propagators, with boundary points in the past or future of a bulk Cauchy slice, span the positive- or negative-energy solutions of the free scalar in AdS$_4$ \cite{Navarro:2026rna}. These are the AdS counterparts of carrollian wavefunctions: they are labeled by a boundary point and diagonalize the boost towards it, with eigenvalue $\Delta$. Their Klein--Gordon inner products on a bulk Cauchy slice are given by CFT$_3$ two-point functions, similarly to \eqref{state overlap}. In the flat limit, the free field then decomposes in a carrollian basis at fixed $\Delta$, and in a plane-wave basis for $\Delta\to\infty$ \cite{Navarro:2026rna}. This is the bulk wavefunction counterpart, for scalar representations, of the massless contraction \eqref{carrollian states bis} and of the massive contraction of Part I \cite{Nguyen:2026rwt}, which the present analysis extends to arbitrary integer spin directly at the level of the Hilbert space.
	
	\paragraph{Inequivalent contractions.}
	As already illustrated by the comparison with \cite{Angelopoulos:1980wg}, contractions of
	representations are not unique, and different schemes can lead to different outcomes
	\cite{Mickelsson:1972fh,Evans:1967zgs,Angelopoulos:1980wg}. Any fixed vector in the carrier space of $\mathcal D(\Delta,s)$, and more generally any descendant at finite level, is annihilated by the translations $\mathcal P_\mu=c\,L_{\mu4}$ in the limit $c\to0$. Non-vanishing flat space momenta are instead carried by descendants at levels of order $n \sim c^{-1}$, in agreement with the scaling of the radial quantum number of mode functions mentioned above. The boosted states \eqref{states tau definition} precisely resum this window, and their norms must be renormalized as in \eqref{tau overlap}. This explains the different outcome of a recent analysis of the flat limit of massless fields in even dimensions, performed both at the level of asymptotic solution spaces and of conformal modules \cite{Bekaert:2026cib}. There, the contraction is implemented on generalized Verma modules, with the primary vector held fixed and descendants at finite level suitably rescaled, without reference to the inner product. In that scheme, the `vev' carrying $\mathcal D(s+1,s)$ contracts to an indecomposable and non-unitary Poincaré module describing charge aspects together with an infinite tower of subleading data, while radiation originates from the `source'. The two results are  not in contradiction: the contraction of \cite{Bekaert:2026cib} acts on vectors at finite descendant level, which necessarily carry vanishing flat space momenta, whereas massless scattering states are carried by descendants at level $n \sim c^{-1}$.
	
	
	\paragraph{Null states, charge aspects and soft sectors.}
	At the unitarity bound, the components $O_J$ with $|J|<s$ become null and independent of retarded time, see \eqref{SC-limit} and Section~\ref{section 4.3}. In radial quantization, the radial direction in $\mathbb R^3$ is mapped to the time direction of the Lorentzian cylinder after Wick rotation. The component $n^iO_{i\,i_2\dots i_s}$ thus corresponds to the time components of the conserved current. Two other analyses point to the very same components. In \cite{Bekaert:2026cib}, the components of spin $s'<s$ of the conserved current become the charge aspect multiplet, which includes charge, mass and angular momentum aspects, and which is annihilated by translations in the flat limit. In \cite{Hijano:2020szl}, the Ward identities of the boundary current, controlled by its time component, give rise to the soft photon theorem. This suggests that the null lower-spin sector found here is the remnant of the soft and charge sectors of flat space gauge theories. Since the latter are not described by normalizable states, they are not expected to appear in the contracted Hilbert space itself.\footnote{A genuine Hilbert space can be associated to Goldstone modes of spontaneously broken asymptotic symmetries \cite{Agrawal:2025bsy}. These representations have strictly zero momentum, i.e., they carry UIRs of the Lorentz group $\operatorname{SO}(1,3)$.} It would be interesting to make this relation precise.
	
	\paragraph{Spinor-helicity formalism in AdS$_4$.}
	A different route to the flat limit of massless fields in AdS$_4$ is provided by the spinor-helicity formalism of Nagaraj and Ponomarev \cite{Nagaraj:2018nxq,Nagaraj:2019zmk,Nagaraj:2020sji}. There, the massless representations are realized on $\mathfrak{sl}(2,\mathbb C)$ spinor variables, such that the representation space splits into sectors of definite helicity $J=\pm s$. AdS counterparts of plane waves are then constructed, and they reduce to flat space plane waves as $\ell\to\infty$. This construction is adapted from the outset to a momentum-like basis and restricted to massless fields. By contrast, the contraction presented here applies to arbitrary $\Delta\geq s+1$ and lands naturally on the carrollian basis. It would be interesting to relate the two descriptions explicitly. 
	
	\paragraph{Towards massless scattering amplitudes.}
	In Part I of this series \cite{Nguyen:2026rwt}, we suggested that the Hilbert space contraction of $\mathcal D(\Delta,s)$ to massive particles directly translates into a prescription for extracting massive scattering amplitudes from conformal correlators. The results presented here suggest an analogous prescription for massless particles:
	\begin{enumerate}
		\item Start from conformal correlators in $\mathbb R^3$, with operator insertions inside and outside the unit ball at radial positions $r_i=1\mp c\,\tau_i$, i.e., within a vanishingly small neighborhood of the unit sphere;
		\item Rescale each operator insertion by $c^{\Delta_i}$, project onto definite helicities using the adapted triad of Section~\ref{section 3.3}, and take the limit $c\to0$ at fixed $(\tau_i,\vec n_i)$ and fixed scaling dimensions $\Delta_i$;
		\item Implement the Wick rotation $\tau_i=-iu_i+0^+$, resulting in carrollian amplitudes, i.e., massless S-matrix elements expressed in the carrollian basis, related to momentum space amplitudes through the modified Mellin transform \eqref{carrollian states bis}.
	\end{enumerate}
	The inner product \eqref{IP-helicity} provides the simplest instance of this procedure, corresponding to the trivial $1\to1$ scattering process. As in the massive case, the limit taken in step 2 requires appropriate rescalings by powers of $c$ and is to be understood in the sense of distributions.
	
	\begin{figure}[H]
		\centering
		\begin{tikzpicture}[scale=1.5,>={Stealth}]
			\fill[blue!7] (-1.65,-1.55) rectangle (1.65,1.55);
			\node[anchor=north west] at (-1.65,1.55) {\footnotesize $\mathbb{R}^3$};
			\fill[orange!18] (0,0) circle (1);
			\fill[red!55,opacity=0.55,even odd rule] (0,0) circle (1) (0,0) circle (0.9);
			\fill[blue!55,opacity=0.55,even odd rule] (0,0) circle (1.1) (0,0) circle (1);
			\draw[thick,orange!50!red] (0,0) circle (1);
			\draw[orange!50!red,dashed] (-1,0) arc (180:360:1 and 0.25);
			\draw[orange!50!red,densely dotted] (1,0) arc (0:180:1 and 0.25);
			\fill (-0.539,-0.770) circle (0.035) node[above right=-1pt] {\scriptsize 1};
			\fill ( 0.539,-0.770) circle (0.035) node[above left=-1pt] {\scriptsize 2};
			\fill ( 0.602, 0.860) circle (0.035) node[above right=-1pt] {\scriptsize 3};
			\fill (-0.602, 0.860) circle (0.035) node[above left=-1pt] {\scriptsize 4};
			\node at (0,0.45) {\scriptsize $x<1$};
			\node at (1.15,-1.3) {\scriptsize $x>1$};
			\draw[->,thick] (1.85,0) -- (3.05,0); 
			\begin{scope}[shift={(4.45,0)},scale=1.3]
				\fill[gray!8] (-0.7,-1) rectangle (0.7,1);
				\draw[thick] (-0.7,-1) -- (-0.7,1);
				\draw[thick] (0.7,-1) -- (0.7,1);
				\fill[orange!18] plot[domain=-0.7:0.7,samples=40] (\x,{-0.45-1.115*(sqrt(0.0625+\x*\x)-0.25)}) -- cycle;
				\draw[thick,orange!50!red] plot[domain=-0.7:0.7,samples=40] (\x,{-0.45-1.115*(sqrt(0.0625+\x*\x)-0.25)});
				\fill[blue!12] plot[domain=-0.7:0.7,samples=40] (\x,{0.45+1.115*(sqrt(0.0625+\x*\x)-0.25)}) -- cycle;
				\draw[thick,blue!60!black] plot[domain=-0.7:0.7,samples=40] (\x,{0.45+1.115*(sqrt(0.0625+\x*\x)-0.25)});
				\fill[red!55,opacity=0.55]
				(-0.5,-0.795) arc (180:360:0.5 and 0.107)
				-- plot[domain=0.5:0.7,samples=10] (\x,{-0.45-1.115*(sqrt(0.0625+\x*\x)-0.25)})
				arc (360:180:0.7 and 0.15)
				-- plot[domain=-0.7:-0.5,samples=10] (\x,{-0.45-1.115*(sqrt(0.0625+\x*\x)-0.25)}) -- cycle;
				\fill[blue!55,opacity=0.55]
				(-0.5,0.795) arc (180:360:0.5 and 0.107)
				-- plot[domain=0.5:0.7,samples=10] (\x,{0.45+1.115*(sqrt(0.0625+\x*\x)-0.25)})
				arc (360:180:0.7 and 0.15)
				-- plot[domain=-0.7:-0.5,samples=10] (\x,{0.45+1.115*(sqrt(0.0625+\x*\x)-0.25)}) -- cycle;
				\draw[very thick,orange!50!red] (-0.7,-1) arc (180:360:0.7 and 0.15);
				\draw[orange!50!red,dashed] (0.7,-1) arc (0:180:0.7 and 0.15);
				\draw[very thick,blue!60!black] (-0.7,1) arc (180:360:0.7 and 0.15);
				\draw[blue!80!black,dashed,thick] (0.7,1) arc (0:180:0.7 and 0.15);
				\fill (-0.344,-1.006) circle (0.03) node[left=0.5pt] {\scriptsize 1};
				\fill ( 0.344,-1.006) circle (0.03) node[right=0.5pt] {\scriptsize 2};
				\fill ( 0.344, 0.794) circle (0.03) node[left=0.5pt] {\scriptsize 3};
				\fill (-0.344, 0.794) circle (0.03) node[right=0.5pt] {\scriptsize 4};
				\draw[<->] (0.95,-1) -- (0.95,1) node[midway,right] {\scriptsize $\pi$};
				\node[left] at (-0.74,-1) {\scriptsize $\tau_{\text{AdS}}=-\tfrac{\pi}{2}$};
				\node[left] at (-0.74,1) {\scriptsize $\tau_{\text{AdS}}=+\tfrac{\pi}{2}$};
			\end{scope}
		\end{tikzpicture}
		\caption{Illustration of the prescription for deriving massless scattering amplitudes from
			conformal correlators. Left: operators are inserted within the thin shells $r_i=1\mp c\,\tau_i$
			enclosing the unit sphere. Right: the interior and exterior of the unit ball, where in and out
			states are prepared, correspond to Euclidean caps glued onto the Lorentzian cylinder at the global
			time slices $\tau_{\text{AdS}}=\mp\tfrac{\pi}{2}$. Through the conformal map
			\eqref{momentum parametrisation}, the two caps may be replaced by the two sheets of the
			pseudo-momentum hyperboloid of Figure~\ref{fig:ball to hyperboloid}, whose asymptotia correspond to the infinite boost limit.}
		\label{fig:massless prescription}
	\end{figure}
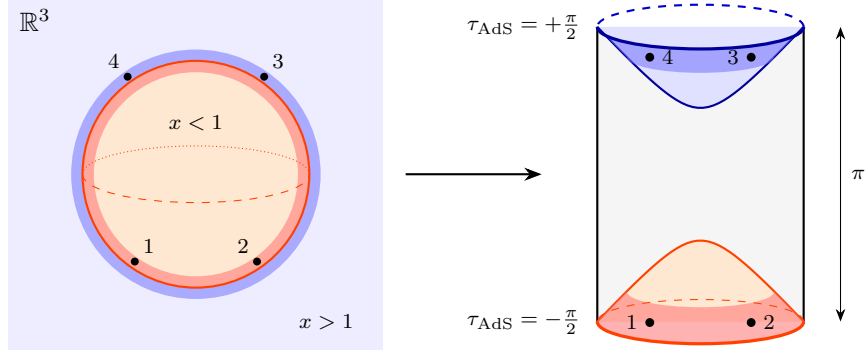
	
	Prescriptions for extracting massless scattering amplitudes from conformal correlators at fixed $\Delta$ have been formulated in Mellin space \cite{Penedones:2010ue,Fitzpatrick:2011ia,Fitzpatrick:2011dm,Fitzpatrick:2011hu} and in position space \cite{Fitzpatrick:2011jn,Fitzpatrick:2011hu,Hijano:2019qmi,Hijano:2020szl,Li:2021snj,Gary:2009ae}, where they are expressed in terms of boundary operators smeared near the global times $\tau_{\text{AdS}}=\pm\tfrac{\pi}{2}$. More recently, they have also been formulated in terms of carrollian amplitudes, with boundary insertions at $\tau_{\text{AdS}}=\pm\tfrac{\pi}{2}+\tfrac{u}{\ell}$ \cite{Bagchi:2023fbj,Alday:2024yyj,Lipstein:2025jfj,deGioia:2022fcn,deGioia:2024yne}. The prescription suggested here is most closely related to the latter, although the boundary insertions are kept Euclidean until the very last step, where a Lorentzian picture emerges only after implementing the Wick rotation $\tau=-i u+0^+$,  whose purpose is to make the inverse transform \eqref{inverse transform bis} run along the real axis. See Figure~\ref{fig:massless prescription}. In the flat limit, the two sheets of the pseudo-momentum hyperboloid naturally correspond to past and future timelike infinity $i^\mp$ where massive scattering states are encoded \cite{Have:2024dff}.
	The infinite boost limit pushes the states towards the conformal boundary of these hyperboloids,
	where timelike infinity meets null infinity, allowing to trade massive particles for
	massless particles.

	The proposed prescription is directly rooted in the Hilbert space contraction described in this work, including the normalization of the external states. It also naturally accommodates spinning operators of arbitrary dimension. Above the unitarity bound, it predicts that correlators of non-conserved spinning operators give rise to amplitudes involving the lower-spin massless particles of \eqref{result generic}, in addition to those of maximal spin. In the spirit of the equivalence theorem, the former should be interpreted as amplitudes of the would-be Goldstone particles associated with the longitudinal modes. At the unitarity bound, only the helicities $J=\pm s$ survive, and conserved currents give rise to amplitudes of massless gauge fields. This is particularly relevant for concrete realizations of AdS/CFT correspondences, whose spectra of scaling dimensions cannot be tuned at will, and which always contain short multiplets dual to the graviton and other gauge fields \cite{Aharony:1999ti}. Applying this prescription to explicit holographic correlators, and relating it to other similar approaches \cite{Bagchi:2023fbj,Alday:2024yyj,Kraus:2024gso,Lipstein:2025jfj}, is left for future work.

	\section*{Acknowledgments}
	
	We thank Fernando Alday, Dionysios Anninos and Robinson Mancilla for stimulating discussions. This work was supported by a postdoctoral research fellowship of the F.R.S.-FNRS (Belgium).
	
	\appendix
	
	\section{Proof of the spin decomposition}
	\label{appendix general spin}
	Every generator $G$ of the Lorentz subalgebra $\mathfrak{so}(1,3)$ acts as
	\begin{equation}\label{genform}
		[G,O_I] = S_G^{(s)} O_I + \mathcal D_G\, O_I\,,
	\end{equation}
	with an orbital part given in terms of a differential operator
	\begin{equation}
		\mathcal D_G = \xi_G^a(y) \partial_a + \chi_G(t,y) \partial_t + w_G(y)\,,
	\end{equation}
	and an intrinsic spin part given by a $y$-dependent combination of spin matrices,
	\begin{equation}
		\label{SG}
		S_G^{(s)} = c_G^{ij}(y) [L_{ij}]^{(s)}\,.
	\end{equation}
	
	Let us denote by $R\indices{_A^i}$ the matrix with components
	\begin{equation}
		\label{R components}
		R\indices{_a^i}=e^i_a\,, \qquad R\indices{_3^i}=n^i\,,
	\end{equation}
	which is orthogonal on account of the completeness relation \eqref{completeness triad}
	\begin{equation}
		R^T R=1\,.
	\end{equation}
	The adapted frame components of the field $O_I$ are given by
	\begin{equation}
		O_{A_1\dots\, A_s}=R\indices{_{A_1}^{i_1}}\dots R\indices{_{A_s}^{i_s}}\, O_{i_1 \dots\, i_s}\,.
	\end{equation}
	A symmetry generator then acts as
	\begin{equation}
		\label{A10}
		[G,O_{A_1\dots\, A_s}]=R\indices{_{A_1}^{i_1}}\dots R\indices{_{A_s}^{i_s}} \big(S_G^{(s)} O
		+ \mathcal D_G  O\big)_{i_1\dots\, i_s}\,.
	\end{equation}
	For the spin term, we use \eqref{SG} and the Leibniz rule \eqref{Leibniz rule} to write
	\begin{equation}
		R\indices{_{A_1}^{i_1}}\dots R\indices{_{A_s}^{i_s}} S_G^{(s)} O_{i_1\dots\, i_s}=\sum_{r=1}^s (R S_G^{(1)} R^T)\indices{_{A_r}^B}\, O_{A_1\dots\, B\, \dots\, A_s}\,.
	\end{equation}
	For the orbital term, we pass the rotation matrices through the derivative operator, yielding
	\begin{equation}
		R\indices{_{A_1}^{i_1}}\dots R\indices{_{A_s}^{i_s}} \mathcal{D}_G O_{i_1\dots\, i_s}=\mathcal{D}_G O_{A_1\dots\, A_s}-\sum_{r=1}^s (\xi^a_G \partial_a R\, R^T)\indices{_{A_r}^B}\, O_{A_1\dots\, B\, \dots\, A_s}\,.
	\end{equation}
	Hence, the symmetry action \eqref{A10} reduces to
	\begin{equation}
		[G,O_{A_1\dots\, A_s}]=\mathcal{D}_G O_{A_1\dots\, A_s}+\sum_{r=1}^s (\sigma_G)\indices{_{A_r}^B}\, O_{A_1\dots\, B\, \dots\, A_s}\,,
	\end{equation}
	in terms of the matrix
	\begin{equation}
		\sigma_G=(R S_G^{(1)}-\xi^a_G \partial_a R) R^T\,.
	\end{equation}
	The determination of the action on a rank-$s$ tensor has been reduced to the determination of the matrix $\sigma_G$ describing the action on the vector representation. We now proceed to the explicit evaluation of this matrix for all $\mathfrak{so}(1,3)$ generators of the carrollian basis. The remaining abelian generators $H, K, B_a$ carry no spin part and therefore do not require treatment.
	
	\paragraph{Carroll dilation.} For the generator $D$, the spin operator and orbital vector field are
	\begin{equation}
		S_{D}^{(1)}=-n^j [L_{3j}]\,, \qquad \xi_{D}^a=-i y^a\,,
	\end{equation}
	such that
	\begin{equation}
		[\sigma_{D}]\indices{_A^B}\, O_B=\left(-R\indices{_A^i}n^j[L_{3j}]+iy^a \partial_a R\indices{_A^i}\right) O_i\,.
	\end{equation}
	Using \eqref{R components} and the explicit expressions for the triad and the vector representation $[L_{ij}]$, one finds
	\begin{equation}
		\sigma_{D}=0\,.
	\end{equation}
	
	\paragraph{Carroll translations.} For the generator $P_a$, we have
	\begin{equation}
		S_{P_a}^{(1)}=n^j[L_{aj}]+[L_{a3}]\,, \qquad \xi_{P_a}^b=-i \delta^b_a\,,
	\end{equation}
	such that
	\begin{equation}
		[\sigma_{P_a}]\indices{_A^B}\, O_B=\left(R\indices{_A^i}(n^j[L_{aj}]+[L_{a3}])+i \partial_a R\indices{_A^i}\right) O_i\,.
	\end{equation}
	Explicit evaluation also yields
	\begin{equation}
		\sigma_{P_a}=0\,.
	\end{equation}
	
	\paragraph{Carroll special conformal transformations.}  For the generator $K_a$, we have
	\begin{equation}
		S_{K_a}^{(1)}=n^j[L_{aj}]-[L_{a3}]\,, \qquad \xi_{K_a}^b=i(2y_a y^b-y^2 \delta^b_a)\,,
	\end{equation}
	such that
	\begin{equation}
		[\sigma_{K_a}]\indices{_A^B}\, O_B=\left(R\indices{_A^i}(n^j[L_{aj}]-[L_{a3}])-i (2y_a y^b \partial_b-y^2 \partial_a) R\indices{_A^i}\right) O_i\,.
	\end{equation}
	Explicit evaluation also yields
	\begin{equation}
		[\sigma_{K_a}]\indices{_c^d}=2y^b [\Sigma_{ab}]\indices{_c^d}\,, \qquad [\sigma_{K_a}]\indices{_3^c}=[\sigma_{ K_a}]\indices{_c^3}=[\sigma_{K_a}]\indices{_3^3}=0\,,
	\end{equation}
	where $[\Sigma_{ab}]$ is the vector representation of $\operatorname{SO}(2)$ generators,
	\begin{equation}
		[\Sigma_{ab}]\indices{_c^d} = i\,(\delta_{ac}\delta_b^d - \delta_{bc}\delta_a^d)\,.
	\end{equation}
	
	\paragraph{Carroll rotations.}  For the generator $L_{ab}$, we have
	\begin{equation}
		S_{L_{ab}}^{(1)}=[L_{ab}]\,, \qquad \xi_{L_{ab}}^c=i(y_a \delta^c_b-y_b \delta^c_a)\,,
	\end{equation}
	such that
	\begin{equation}
		[\sigma_{L_{ab}}]\indices{_A^B}\, O_B=\left(R\indices{_A^i}[L_{ab}]-i (y_a \partial_b-y_b \partial_a) R\indices{_A^i}\right) O_i\,.
	\end{equation}
	Explicit evaluation also yields
	\begin{equation}
		[\sigma_{L_{ab}}]\indices{_c^d}=[\Sigma_{ab}]\indices{_c^d}\,, \qquad [\sigma_{L_{ab}}]\indices{_3^c}=[\sigma_{ L_{ab}}]\indices{_c^3}=[\sigma_{L_{ab}}]\indices{_3^3}=0\,.
	\end{equation}
	
	\bibliography{bibl}
	\bibliographystyle{JHEP}  
\end{document}